\documentclass[a4paper,12pt]{article}
\usepackage{jheppub}
\usepackage[toc,page]{appendix}
\usepackage{hyperref}
\usepackage{color}
\usepackage{graphicx,float}
\usepackage{ulem}
\usepackage{color}
\usepackage{overpic}
\usepackage{subcaption}
\usepackage{bm}

\usepackage[mathscr]{euscript}
\usepackage[usenames,dvipsnames]{xcolor}
\usepackage{amsmath,amsfonts,amssymb,verbatim,float,mathtools}
\usepackage{physics}
\usepackage{resizegather}
\usepackage{tikz}
\usetikzlibrary{shapes.misc}
 \tikzset{cross/.style={cross out, draw=black, minimum size=2*(#1-\pgflinewidth), inner sep=0pt, outer sep=0pt},
cross/.default={4pt}}

\usepackage{empheq}

\usepackage[most]{tcolorbox}
\tcbset{myformula/.style={colback=white, %yellow!10!white,
    colframe=black, %red!50!black,
    top=4pt,bottom=4pt,left=0pt,right=4pt,
    boxsep=0pt,
    arc=0pt,
    outer arc=0pt,
    fit algorithm=hybrid*
}}
\newcommand\fiteq[1]{%
  \sbox{\mybox}{$\displaystyle#1$}%
  \ifdim\wd\mybox>.85\textwidth\resizebox{.85\textwidth}{!}{\usebox{\mybox}}%
  \else\usebox{\mybox}\fi%
}
\newsavebox{\mybox}

\newtcolorbox{equationframe}{
math
}

\definecolor{orange}{rgb}{1,0.5,0}
\definecolor{col1}{RGB}{153, 52, 121}
\definecolor{dgreen}{rgb}{0,0.55,0}
\definecolor{pink}{rgb}{1,0.08,0.58}

\newcommand{\la}{\langle}
\newcommand{\ra}{\rangle}
\newcommand{\p}{\partial}

\newcommand{\rar}{{\rightarrow}}

\newcommand{\bs}{\begin{split}}
\def\sess\end{split}

\definecolor{col1}{RGB}{153, 52, 121}

\graphicspath{{folder/}{images/}}

\begin{document}

\preprint{NORDITA 2023-068}

\title{Holographic timelike superconductor}

\author[a]{Alexander Krikun\footnote{https://orcid.org/0000-0001-8789-8703}}

\affiliation[a]{Nordita KTH Royal Institute of Technology and Stockholm University \\
Hannes Alfv\/ens v\"ag 12, SE-106 91 Stockholm, Sweden}

\author[a,b]{and Uriel Elinos\footnote{Temporary at Nordita as a visiting PhD fellow}}

\affiliation[b]{
Departamento de F\/isica, Universidad Aut\/onoma Metropolitana—Iztapalapa, \\
Avenida Ferrocarril San Rafael Atlixco 186, C.P. 09340, Ciudad de México, México
%Metropolitan Autonomous University (UAM), Mexico
}

% e-mail addresses: one for each author, in the same order as the authors

\emailAdd{krikun@nordita.org}
\emailAdd{elinosur@gmail.com}

\abstract{
We explore the state of matter characterized by the charged timelike vector order parameter. We employ holographic duality in order to construct such a state and study its thermoelectric transport, fermionic spectral function and the sign of the Meissner effect. We uncover the unusual features of this ``timelike superconductor'': the absence of the gap in the fermionic spectrum and co-existence of Drude peak and supercurrent in the AC transport, which are reminiscent to those of time-reversal-odd and gapless superconductors, correspondingly. We show that this state is dynamically stable and thermodynamically at least metastable. Along the way we develop the holographic model of the charged vector field, which acquires mass due to a variant of the Stueckelberg mechanism with the adjoint Higgs field.
}

\maketitle

\section{Introduction}

% As it has been found in \cite{Arias:2016nww}, this model allows for a spontaneous condensation of the temporal component of the Proca field. This doesn't break isotropy of the state, therefore this phase has been identified with and S-wave superconductor. However, as we elaborate in much more detail throughout this study, this condensate breaks time reversal symmetry and therefore represents (contrary to what has been attributed in \cite{Arias:2016nww} the quite exotic state: time-reversal odd s-wave superconductor. 

Time-reversal-odd superconductivity is a state in Condensed matter, which has been shown to be possible theoretically, but so far has not been observed experimentally \cite{linder2017odd}. Fundamentally, the state is characterized by the order parameter which is charged under electromagnetic $U(1)$ gauge field, hence ``superconductivity'' (SC), and is odd under time (T) reversal. This is unlike the usual superconductivity which is T-even. The fact that the order parameter is T-odd can lead to several interesting consequences \cite{balatsky1994properties}. In particular, it may not form the superconducting gap in the fermion spectrum and may not affect the DC transport. These features make it hard to observe this state
% in phenomenology and therefore T-odd superconducting order is one of the candidates for the ``hidden order'', which may govern the pseudogap state of various strongly correlated materials, like cuprate superconductors and heavy fermion systems. 
and there is currently no theoretical agreement on what are its specific phenomenological signatures, since various observables seem to depend crucially on the details of the fermionic gap function used in conventional condensed matter calculations \cite{sukhachov2019spectroscopic}.

The holographic duality \cite{Zaanen:2015oix} can be used as a phenomenological tool for studying various ordered states which is not relying on the details of the mechanism of condensation and therefore allows one to make general statements about the features of a given state, provided the symmetries of the order parameter are known. This includes i.e. the models for p-wave superconductor \cite{Gubser:2008wv,Cai:2013aca}, d-wave superconductor \cite{Benini:2010pr}, charge density wave \cite{Donos:2013gda} (\cite{Baggioli:2022pyb} for review). Unlike the effective Ginzburg-Landau approach, however, in holographic model a whole range of the observables can be consistently computed, which provides a useful insight on the phenomenological features of a given condensed state.

In this work we will use the holographic duality in order to study a state, which belongs to the class of T-odd superconductors and is characterized by the 4-vector order parameter, which condenses in the timelike direction.    
From the symmetry perspective, this order parameter is the same as the one describing the more conventional $p$-wave superconductor, where the vector condenses in the spatial direction instead. The novel aspect, which we introduce here is the very possibility for this vector to be directed in time. It is therefore a crucial assumption on the dual condensed matter system, that there is an underlying Lorentz symmetry. The systems are known, however, where this symmetry is emergent at low energy: the Dirac materials including graphene.

In fact the timelike SC state, which we are going to explore, has been already observed in the holographic model with a complex Proca field in the bulk \cite{Cai:2013pda,Cai:2013aca}. It has been shown already in \cite{Arias:2016nww}, that besides the condensed phase with P-wave order, this model can realize a novel isotropic state, where the temporal component of the massive vector field condenses in the bulk. In \cite{Arias:2016nww}, however, this state has not been in the main focus and its T-odd features have been overlooked. Moreover the timelike state was shown to be dynamically unstable towards formation of the persistent current.

Here we present a comprehensive study of the timelike superconductor. We start in Sec.\ref{sec:model} by presenting the improved holographic model, which describes the massive charged vector field relying on the (explicitely borken) adjoint Higgs mechanism. We show that in our model the state under consideration is indeed stable. We then proceed in Sec.\ref{sec:transport} with the study of AC thermoelectric transport features and the analysis of the dynamical stability. In Sec.\,\ref{sec:fermions} we evaluate the fermionic spectral function and Sec.\ref{sec:meissner} is devoted to the study of the Meissner effect. We conclude in Sec.\ref{sec:conclusion}, providing more details on the equations of motion, fermionic coupling terms, perturbative treatment of the Meissner effect and the numercial treatment in the Appendices.

\section{\label{sec:model}The model and the condensed phase}

The holographic models for P-wave superconductivity have been studied earlier and come in roughly two classes: the non-Abelian gauge field models \cite{Gubser:2008wv,Ammon:2009xh} and Abelian charged Proca models \cite{Cai:2013pda, Cai:2013aca,Arias:2016nww}. In the former the spontaneous condensation of one of the gauge field components breaks simultaneously the symmetry of rotations (P-wave) and the non-Abelian symmetry, leaving a residual $U(1)$ behind, which is then interpreted as the dual to electromagnetic charge on the boundary. The condensed $SU(2)$ bosons are therefore naturally charged under this residual $U(1)$ and represent P-wave superconducting order parameter. In the latter class of models one starts by postulating the existence of the charged vector order parameter and develops the holographic model with a complex ($U(1)$ charged) vector field, which does not correspond to any gauge symmetry and generally has a nonzero mass -- the charged Proca field. 
On one hand, the extra non-Abelian gauge symmetry of the former model has no motivation from the phenomenology point of view, but restricts considerably the dynamics of the model. The nonzero mass of the vector field is in this regard the advantage of the latter model, which allows for a more complex phase structure. However, as we discuss in more detail in the Appendix \ref{app:charged_Proca}, while the theory of neutral Proca field is consistent at the classical level, the complex (charged) Proca field is prone to the occasional inconsistencies depending on the background U(1) gauge field, which makes this model ill-defined in the general background. In this work we will combine the two approaches and get rid of these disadvantages. 

We write down the holographic model in 3+1 bulk space of curvature $R$, with $U(1)$ gauge field $A_\mu$, a complex vector field $B_\mu$, a complex scalar $\Phi$ and a neutral scalar $\phi$. The action reads
\begin{align}
\label{equ:Proca_part}
S = \int d^4 x \sqrt{-g} \bigg(& R + 6 - \frac{1}{4} F_{\mu \nu} F^{\mu \nu} - \frac{1}{2} W_{\mu \nu} \bar{W}^{\mu \nu} - q_B^2 \phi^2 B_{\mu} \bar{B}^{\mu}  \\
\notag
&  + i q_B {F}^{\mu \nu} B_{[\mu} \bar{B}_{\nu]} + \frac{1}{2} q_B^2 \left(B^2 \bar{B}^2 - (B_{\mu} \bar{B}^{\mu})^2 \right)\\
\notag
& - \frac{1}{2} (\p_{\mu} \phi)^2  - \frac{1}{2} m^2 \phi^2 
- D_{\mu} \Phi D^{\mu} \bar{\Phi}  - m^2 \Phi \bar{\Phi}  \\
\notag
&+ \frac{1}{2} q_B^2 \left[B^2 \bar{\Phi}^2 + \bar{B}_{\mu}^2 \Phi^2 - 2 \bar{B}_{\mu} B^{\mu} \bar{\Phi} \Phi \right] \\
\notag
&+ i q_B \phi ( \bar{B}^{\mu} D_{\mu} \Phi - B^{\mu}  D_{\mu} \bar{\Phi} ) + i q_B \p_{\mu} \phi (B^{\mu} \bar{\Phi} - \bar{B}^{\mu} \Phi)
\bigg)
\end{align}
\begin{gather}
\notag
W_{\mu \nu} = D_\mu B_\nu - D_\nu B_\mu, \qquad F_{\mu \nu} = \p_{\mu} A_{\nu} - \p_{\nu} A_{\mu} \\ 
\notag
B^2 \equiv B_{\mu} B^{\mu}, \qquad \bar{B}^2 \equiv \bar{B}_{\mu} \bar{B}^{\mu} \\
\notag
 \qquad D_{\mu} B_{\nu} = \p_{\mu} - i q_B A_{\mu}  B_{\nu}, \qquad D_{\mu} \Phi = \p_{\mu} - i q_B A_{\mu}  \Phi,
\end{gather}
In the first line one immediately recognizes the complex Proca action considered in i.e. \cite{Cai:2013aca,Arias:2016nww}, where we set the coefficient of magnetic moment term $FB\bar{B}$ to 1 ($\gamma$ in eq.(1) of \cite{Cai:2013aca} and absent in \cite{Arias:2016nww}). In this way one can interpret $A_\mu$ as dual to the $U(1)$ charge density and chemical potential and $B_{\mu}$ as dual to the p-wave superconducting order parameter. 
We also promote the Proca mass to the dynamical $\phi$-field. We will keep control over the vector mass by setting the boundary value of $\phi$. 
Besides this similarity, we introduce a set of extra nonlinear self-interaction terms for $B$-fields and the interactions between $B_\mu$, $\phi$ and the extra complex scalar $\Phi$. 

The structure of the extra interaction terms becomes clear once we notice that the action above is algebraically equivalent to the action of the non-Abelian $SU(2)$ model in the spirit of \cite{Gubser:2008wv} with an extra Higgs field in the adjoin representation
\begin{equation}
\label{equ:SU2_adj_Higgs_action}
S = \int d^4 x - \frac{1}{2} \tr \mathcal{G}_{\mu \nu}^{\dag} \mathcal{G}^{\mu \nu} - \tr (D_{\mu} \mathbf{\Phi}^{\dag} D^{\mu} \mathbf{\Phi}) - m^2 \tr(\mathbf{\Phi}^{\dag}\mathbf{\Phi}),
\end{equation}
\begin{align}
\notag
\mathcal{G}_{\mu \nu} &= \p_\mu \mathcal{B}_{\nu} - \p_\nu \mathcal{B}_{\mu} - i q_B [\mathcal{B}_\mu, \mathcal{B}_\nu], 
&
\mathcal{B}_{\mu} &= B_{\mu} \tau^+ + \bar{B}_{\mu} \tau^- + A_{\mu} \tau^3 \\
\notag
D_{\mu} \mathbf{\Phi} &= \p_\mu \mathbf{\Phi} - i q_B [\mathcal{B}_{\mu}, \mathbf{\Phi}],
&
\mathbf{\Phi} &=  \Phi \tau^+ + \bar{\Phi} \tau^- + \phi \tau^3,
\end{align}
with $SU(2)$ generators chosen as the rising/lowering combinations of the Pauli matrices
\begin{equation}
\tau^+ = \frac{1}{2 \sqrt{2}}(\sigma_1 + i \sigma_2), \qquad 
\tau^- = \frac{1}{2 \sqrt{2}}(\sigma_1 - i \sigma_2), \qquad 
\tau^3 = \frac{1}{2} \sigma_3.
\end{equation}

Written in the form \eqref{equ:Proca_part}, the action exhibits the explicit U(1) gauge invariance, but thanks to all the nonlinear terms it is also invariant under the other remaining SU(2) gauge transformations evident in \eqref{equ:SU2_adj_Higgs_action}. A particularly interesting for us is:
\begin{gather}
\label{equ:Bz_gauge_transform}
B_{z} \rar  B_z  + \frac{1}{\sqrt{2}} \frac{1}{q_B} \p_z \alpha, \\
\notag
\begin{pmatrix}
B_{t} \\ A_t
\end{pmatrix}
\rar 
\begin{pmatrix}
\cos(\alpha) & -\frac{1}{\sqrt{2}}\sin(\alpha) \\
\sqrt{2} \sin(\alpha) & \cos(\alpha) 
\end{pmatrix}
\begin{pmatrix}
B_{t} \\ A_t
\end{pmatrix}, 
\quad
\begin{pmatrix}
\Phi \\ \phi
\end{pmatrix}
\rar 
\begin{pmatrix}
\cos(\alpha) & -\frac{1}{\sqrt{2}}\sin(\alpha) \\
\sqrt{2} \sin(\alpha) & \cos(\alpha) 
\end{pmatrix}
\begin{pmatrix}
\Phi \\ \phi
\end{pmatrix}.
\end{gather}

The extra gauge freedom allows to consistently describe all the degrees of freedom of the charged massive vector fields, fixing the drawback of the plain complex Proca action of \cite{Cai:2013aca}. By setting the boundary value of the scalar field we break the $SU(2)$ symmetry explicitly and provide the vector bosons with a mass via a Stueckelberg/Higgs mechanism. Similarly to the electroweak $W$-bosons, the $B$-fields in \eqref{equ:Proca_part} acquire a third polarization due to the captured $\Phi$ scalar. In this way the number of the propagating degrees of freedom in this setup corresponds to the massive spin-1 charged particle.

We also note that by means of the residual gauge symmetry \eqref{equ:Bz_gauge_transform} we can get rid of the $\Phi$ field and find a solution in the ansatz
\begin{gather}
\label{equ:Bz_gauge}
\mbox{\textit{``Physical'' gauge:}}\quad B = B_t(z) dt + i B_z(z) dz, \quad A = A_t(z) dt, \quad \Phi = 0 \\
\label{equ:A_ansatz} 
\phi(z)\Big|_{z\rar0} = \frac{3}{4}\frac{1}{q_B^2} \qquad A_t\Big|_{z\rar0} = \mu.
\end{gather}
This corresponds exactly (apart from the extra radial dynamics in $\phi(z)$) to the ansatz used in \cite{Arias:2016nww, Cai:2013aca}, where the condensation of the temporal component of the Proca field with $m_B^2 = 3/4$ has been observed and we'll choose the boundary value for the $\phi$-field representing the same asymptotic mass throughout this study. Noteworthy, we set the mass of the adjoint Higgs scalar to $m_{\Phi}^2 = 0$, in order for its leading branch to behave as a constant at $z\rar0$.
Following the interpretation of the holographic model assumed from the Proca perspective \cite{Arias:2016nww}, the chemical potential is dual to the boundary value of the field $A_t$ \textit{in the gauge} \eqref{equ:Bz_gauge}. In the same way we require that the spontaneously condensing $B_\mu$ field has no sources in this ``physical'' gauge \eqref{equ:Bz_gauge}.

The extra gauge symmetry is also helpful when looking for the solutions to the nonlinear equations of motion, where we can use a different ``radial'' gauge $B_z = 0$:
\begin{equation}
\label{equ:radial_gauge}
\mbox{\textit{``Radial'' gauge:}}\quad B = B_t(z) dt, \quad A = A_t(z) dt, \quad \Phi = \Phi(z), \qquad \phi = \phi(z), 
\end{equation}
with a nontrivial dynamics in $\Phi(z)$. As usual in the theories with gauge symmetry, this gauge will allow us to write all the equations of motion for fields $A_t, B_t, \Phi, \phi$ as second order ordinary differential equations and take into account the constraint following from the $B_z$ equation only when dealing with the boundary conditions, see Appendix\,\ref{app:DeTruck}.

It is worth mentioning that the ``extra'' gauge transformation \eqref{equ:Bz_gauge_transform} affects $A_\mu$ and therefore its identification with the electric $U(1)$ boundary current is not gauge invariant. However, we are going to break this symmetry explicitly by introducing the boundary value of the scalar field $\phi$ \textit{in the same gauge} \eqref{equ:Bz_gauge}. Therefore the identification of chemical potential \eqref{equ:A_ansatz} is unambiguous.  
One can view this construction as a softer version of the Stueckelberg trick, where instead of making the Higgs mode completely non-dynamical, we anchor its boundary value.
When working in the ``radial'' gauge, we will need to keep track on how the boundary values of $B_t$ and $A_t$ are affected by the transformation \eqref{equ:Bz_gauge_transform} with $\p_z \alpha = - \sqrt{2} q_B B_z$.

We include a full backreaction of the vector condensation on the geometry, therefore our ansatz for the (isotropic) bulk metric is
\begin{gather}
\label{equ:ds_ansatz}
ds^2 = \frac{1}{z^2}\left[- T(z) f(z) dt^2 + W(z) (dx^2 + dy^2) + \frac{U(z)}{f(z)} dz^2 \right], \\
\notag
f(z) = (1-z)\left(1 + z + z^2 - \bar \mu^2 z^3 / 4 \right)
\end{gather}
This ansatz assumes the background geometry includes a planar black hole and we rescale the radial ($z$) coordinate in such a way that the horizon is located at $z_h = 1$.
The surface gravity at the horizon controls the temperature in the dual field theory, which reads
\begin{equation}
\label{equ:temperature}
T = \frac{12- \mu^2}{16 \pi}.
\end{equation}
Here and in what follows we will consider all the dimensionful parameters in units of $\mu$. 

Finally in order to study meaningful transport properties we introduce a set of extra scalar fields representing the isotropic Q-lattice \cite{Donos:2013eha,Donos:2014uba}. It allows to break the symmetry of translations and obtain finite resistivities in the normal metallic state. The extra action terms introducing Q-lattice are
\begin{equation}
\label{equ:Q-lat}
S_{Q} = \int d^4 x \sqrt{-g} \left( - \p \xi^2 - m_{\xi}^2 \xi^2 - \xi^2 (\p \chi_1^2 + \p \chi_2^2) \right).
\end{equation}
The \textit{axion} fields $\chi_i$ have the specific linear profiles on-shell
\begin{equation}
\label{equ:axions}
\chi_1 = p x, \qquad \chi_2 = p y.
\end{equation}
and break translations in such a way that the coordinate dependence drops out from the equations of motion. This allows for treatment of translation symmetry breaking without introducing partial differential equations. The scalar $\xi$ couples the axions to the geometry and its boundary value $\lambda$ controls the strength of the explicit translational symmetry breaking. In particular, following \cite{Donos:2013eha} we fix the mass of the scalar to $m_{\xi}^2 = -2$, which leads to the near boundary behavior 
\begin{equation}
\label{equ:xi_lambda}
\xi = \xi(z), \qquad \xi \Big|_{z\rar0} = z \lambda + O(z^2).
\end{equation}

Given all these ingredients, we can solve the nonlinear classical equations of motion following from the action \eqref{equ:Proca_part} in presence of the Q-lattice \eqref{equ:Q-lat}, using the isotropic ansatze for the metric \eqref{equ:ds_ansatz}, the Proca vector in the radial gauge \eqref{equ:radial_gauge} and symmetry breaking axions \eqref{equ:axions}, and keeping the fixed values for the temperature \eqref{equ:temperature}, chemical potential \eqref{equ:A_ansatz} and the scale of explicit translational symmetry breaking \eqref{equ:xi_lambda}. We use the ``physical'' boundary values \eqref{equ:A_ansatz} in the radial gauge. Once the solution is obtained, we transform it to the ``physical'' gauge by means of \eqref{equ:Bz_gauge_transform} and check \textit{a posteriori} that the boundary conditions are not affected by the transformation. We add the details of our numerical treatment of the Proca equations of motion and the Einstein equations in Appendix \ref{app:DeTruck}

\begin{figure}
\includegraphics[width=0.55\linewidth]{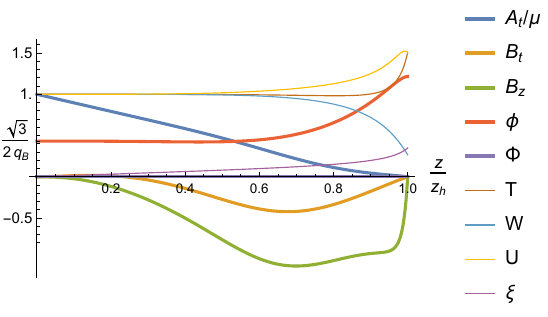}
\includegraphics[width=0.45\linewidth]{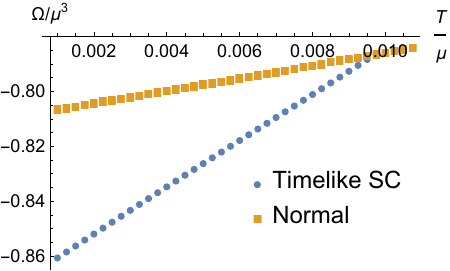}
\caption{\label{fig:Back_solutions}\textbf{Holographic timelike superconductor state.}
\textit{Left panel}: background profiles of the bulk fields in the ``physical'' gauge \eqref{equ:Bz_gauge} with metric ansatz \eqref{equ:ds_ansatz}. This particular solution is evaluated at $T/\mu = 0.005$. \\ 
\textit{Right panel}: The thermodynamic potential of the timelike SC state and normal state. All solution are evaluated for the model parameters \eqref{equ:parameters}. }
\end{figure}

For concreteness, we chose the following values for the model parameters:
\begin{equation}
\label{equ:parameters}
q_B = 2, \qquad \phi(z) \bigg|_{z\rar0} = \sqrt{\frac{3}{4}} q_B^{-1}, \qquad \lambda/\mu= 0.05, \qquad p/\mu = 1/\sqrt{2}.
\end{equation}
We observe a branch of condensed timelike state solutions emerging at temperatures below $T_c \approx 0.0098 \mu$. The branch is characterized by the nonzero profiles of both $B_t$ and $B_z$ components of the Proca field (in physical gauge), as we show on the left panel of Fig.\,\ref{fig:Back_solutions}. Computing the thermodynamic (TD) potential of the new branch of solutions (right panel of Fig.\,\ref{fig:Back_solutions}) we confirm that these are indeed the preferred states as compared to the ``normal'' state with $B_{\mu} = 0$. 

Before we start exploring the features of the newly obtained condensed solution, it is worth mentioning that the same model admits another type of condensation, where the spatial components of the Proca field condense. We were able to find a so-called ``$P+i P$'' state, which is characterized by $B_x = i B_y$ and isotropic geometry\footnote{Interestingly, we could not find a stable ``p-wave'' solution in the model, which would be characterized by the only condensation of $B_x$ and anisotropic geometry.}. For our particular choice of model parameters \eqref{equ:parameters} this branch has a higher critical temperature, $T_c \approx  0.047 \mu$ and correspondingly lower thermodynamic potential at given temperature. Therefore, strictly speaking, the timelike condensed state, which we are focusing on, is not the global minimum of the TD potential and therefore is only metastable. This fact will not affect the features of the timelike state, which we are going to study below. We will also check the dynamical stability of the timelike state. In what follows we will use the ``$P+i P$'' solution as a reference. 

% For the purposes of studying the transport properties and fermionic spectra the global stability of the solutions plays no role. 
% We check, however, that these solutions are dynamically stable, what proves that they realize some local minimum of the free energy indeed.  

\section{\label{sec:transport}Transport properties}
We start exploring the phenomenological features of the novel ``timelike superconductor'' state by studying its transport properties.
In order to study the optical (AC) transport in our holographic model it is enough to consider the time-dependent linear perturbations of the fields around the backgrounds shown on Fig.~\ref{fig:Back_solutions}. 
Being interested in the matrix of the thermoelectric conductivities we  introduce the time dependent perturbative sources for the electric current and energy flux in the stress-energy tensor \cite{Kovtun:2012rj}. These are encoded in the near-boundary asymptotes of the gauge field $A_x(z)$ and metric component $g_{tx}(z)$. 
These perturbations couple at the linear level to several others, which leads us to the set of 6 linear coupled equations of motion on
\begin{equation}
\label{equ:transport_perturbations}
\{\delta g_{tx}, \delta A_x, \delta g_{xz}, \delta B_{x}, \delta \bar B_{x}, \delta \chi_1\} \sim e^{i \omega t}.
\end{equation}
Note that since the background solution is isotropic, we can focus on the $x$-axis without loss of generality.
Once the solutions with the given sources are found (we discuss the methods to solve the perturbation equations in Appendix \ref{app:transport}), we read off the subleading components of the $\delta A_x(z)$ and $\delta g_{tx}(z)$ in order to obtain the two-point functions $\la J^x J^x \ra$, $\la J^x T^{tx} \ra$, $\la T^{tx} J^x \ra$, and $\la T^{tx} T^{tx} \ra$. 

Recalling the definition of the heat current in presence of a chemical potential \cite{Herzog:2009xv}: $Q^x = J^x + \mu T^{t x}$, we can relate the full matrix of AC thermoelectric conductivities \cite{Hartnoll:2009sz} ($\bar \kappa$ is thermal conductivity at zero bias) to the just obtained set of 2-point functions.
\begin{equation}
\label{eq:sigma_matrix}
\begin{pmatrix}
J^x \\ Q^x
\end{pmatrix} = 
\begin{pmatrix}
\sigma & T \alpha \\ T \bar \alpha & T \bar \kappa
\end{pmatrix}
 \begin{pmatrix}
E_x \\ - \frac{\p_x T}{T}
\end{pmatrix}.
\end{equation}

\begin{figure}[t]
\raisebox{-0.5\height}{\includegraphics[width=0.5\linewidth]{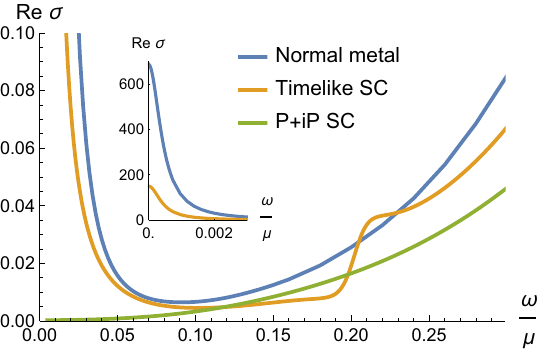}}
\raisebox{-0.\height}{
\begin{minipage}{0.45\linewidth}
\includegraphics[width=1\linewidth]{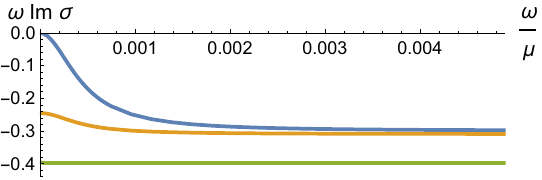}\\
\includegraphics[width=1\linewidth]{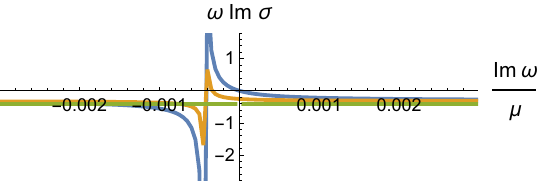}
\end{minipage}
}
\caption{\label{fig:sigma}Electric conductivity for T-even and T-odd SC as well as normal metal
All the states are evaluates at $T/\mu = 0.005$ in the model \eqref{equ:Proca_part} with the model parameters \eqref{equ:parameters}. The normal state is characterized as $B_{\mu} = 0$, the timelike SC state has $B_t, B_z \neq 0$ (see Fig.\,\ref{fig:Back_solutions}, the $P+iP$ state has $B_x = i B_y \neq 0$. 
}
\end{figure}

The results which we obtain for the electric conductivity in the concrete example of the timelike states (yellow curves) with model parameters \eqref{equ:parameters} are shown on Fig.\,\ref{fig:sigma}. For these examples We choose the temperature $T/\mu = 0.005$, which is well below the critical point, i.e. in the developed condensed state. 
For comparison we also plot the AC conductivity for normal metallic state (blue curves) as well as for the
$P+i P$ state, mentioned above (green curves, see Appendix \ref{app:transport} for the details of this calculation). 
The latter, while realizing a combination of two P-wave superconducting orders, is isotropic, has no nodes in th egap function and therefore provides us with an example of the ``conventional'' superconducting phenomenology in the same model at the same external conditions.

In presence of the finite source for translation symmetry breaking $\lambda$, the blue curve exhibits a sharp Drude peak and finite DC value (see the inset on the left panel of  Fig.\,\ref{fig:sigma}). This is expected for a metal with finite resistivity. This peak is a reflection of the pole lying in the lower complex half-plane, which we reveal by computing the imaginary part of the conductivity at purely imaginary frequency (see the lower part of the right panel of  Fig.\,\ref{fig:sigma}). This demonstrates a conventional Drude-like transport 
\begin{equation}
\label{equ:Drude}
\sigma_{\mathrm{metal}} (\omega) \sim \frac{1}{i \omega - \Gamma}, \qquad \Gamma \sim \lambda
\end{equation}
where $\Gamma$ is the momentum relaxation rate due to explicit translation symmetry breaking sources $\lambda$.
Finally, by evaluating the imaginary part of conductivity at small real frequencies we confirm that it vanishes at $\omega = 0$, meaning that there is no pole at $Im(\sigma)$ and therefore there is no superconductivity present in the metallic state.

On the other hand, the ``conventional'' nodeless $P+i P$ state (green curves) has no Drude peak at all in AC conductivity (see the left panel of  Fig.\,\ref{fig:sigma}). This is consistent with the absence of the pole at negative imaginary frequency (lower part of right panel). However, the spectral weight is absorbed in this case in the $\delta$-function in the real part of the conductivity, which follows according to the Kramers-Kronig relations from the existence of a pole in the imaginary part of the conductivity at $\omega =0$, see the top part on the right panel of Fig.\,\ref{fig:sigma} (note that the conductivity is rescaled by $\omega$ for clarity, therefore the simple pole in $Im \sigma$ corresponds to the finite value of $\omega  Im \sigma$ at $\omega  =0$).
\begin{equation}
\label{equ:SC_delta}
\sigma_{\mathrm{superconductor}} \sim \pi \delta(\omega) - i \frac{1}{\omega} 
\end{equation}
This feature describes a phenomenology of the conventional gapped superconductor: the normal transport, characterized by the Drude peak, is gapped and removed, while the superconductivity is mediated by the condensate of charge carriers.  

Finally the timelike state (yellow curves) presents a curious mixture of the metallic and superconducting transport. On one hand, one can clearly observe a Drude peak on the inset in the left panel of Fig.\,\ref{fig:sigma}, which corresponds to the pole at the negative imaginary frequency (the lower part of right panel). On the other hand, the pole in the imaginary conductivity at $\omega=0$ remains, which corresponds to the presence of supercurrent. This phenomenology resembles the class of \textit{gapless superconductors}, which support the supercurrent, but the normal transport is not gapped. 
One would expect such a behavior based on the symmetry properties of the considered state. The time component of the vector order parameter $B_t$ is \textit{time-reversal odd} therefore the order parameter can not manifest itself at zero frequency (due to corresponding odd symmetry in $\omega$) and therefore the AC transport retains a normal component at $\omega = 0$.

However, at finite frequency the order parameter does affect the spectrum and this leads to another notable feature in the AC transport of a timelike SC state: the kink in the spectral density at finite frequency ($\omega/\mu \approx 0.2$ in this case). Comparing to the curves in the normal and $P+iP$ SC states we also observe that unlike conventional superconductor, the timelike superconductor performs a spectral weight transfer to both zero frequency (supercorrent) and higher frequency (the spectral kink). 

Finally, we note that the absence of any poles at positive imaginary frequency (lower part of the right panel of Fig.\,\ref{fig:sigma}) confirms the dynamical stability of the considered timelike superconductor condensed phase.     

\begin{figure}
\raisebox{-0.5\height}{\includegraphics[width=0.5\linewidth]{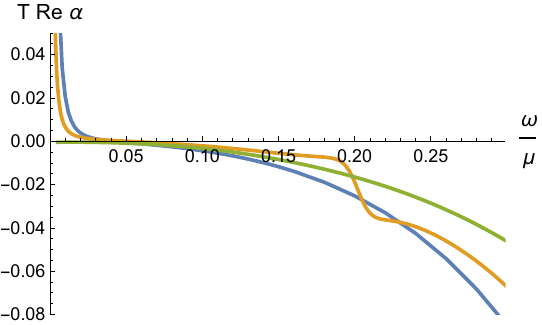}}
\raisebox{-0.5\height}{\includegraphics[width=0.5\linewidth]{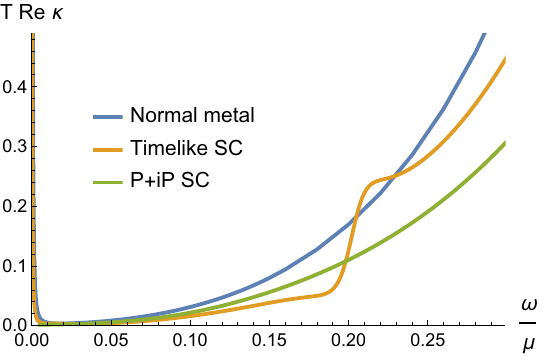}}
\caption{\label{fig:kappa}Thermopower and thermal conductivity for the normal metallic, timelike superconducting and $P+i P$-wave states. The states are the same as the ones on Fig.\,\ref{fig:sigma}}
\end{figure}

For completeness, we show the plots of AC thermopower and thermal conductivity in all three considered states On Fig.\,\ref{fig:kappa}. We have no extra comments here.

\section{\label{sec:fermions}Fermionic spectral function}
In holographic models the fermionic dynamics has a secondary role since due to the presence of the other ``quantum critical'' degrees of freedom in the boundary theory the collective responses, like the transport discussed above, are not related to the fermion propagators \cite{Zaanen:2015oix}. It is however interesting to study the fermionic spectral function for phenomenological purposes, since this is what one would observe when performing i.e. ARPES experiments on the material. 

In order to obtain the fermionic spectral function $\rho = Im Tr \la \bar \Psi \Psi \ra$ one has to evaluate the two-point function of the fermionic operator $\Psi$. In holographic approach this amounts to populating the bulk with the fermionic field, dual to $\Psi$ and solving the corresponding Dirac equation in curved background \cite{Liu:2009dm,Iqbal:2009fd,Cubrovic:2009ye}. In order to take into account the effects of the condensed charged order parameter, the corresponding interaction terms must be included \cite{Faulkner:2009am,Benini:2010pr,Benini:2010qc}. For the vector order parameter, related to P-wave superconductor this has been partially done in \cite{Vegh:2010fc} and, more recently, in \cite{Ghorai:2023wpu}. We note here, that the mass and charge of the bulk fermion and its interaction terms are extra model parameters, which characterize the probe. Therefore the fermionic spectral function, which we obtain for the given state, will generically depend on the particular choices of the probe used. It will be particularly interesting to look for the features which persist for all probes and are therefore characteristic to the groundstate itself. 
We write down the fermionic bulk action:
\begin{equation}
\label{equ:fermion_action}
S_\Psi = - \int d^4 x \sqrt{-g} \left[i \bar{\Psi} (\Gamma^{\underline{\lambda}} e^{\mu}_{\underline{\lambda}} D_{\mu} - m_f) \Psi  + c.c. + \mathcal{L}_{int}\right]
\end{equation}
with covariant derivative
\begin{equation}
\notag
\label{equ:fermion_cov_div}
D_\mu = \p_\mu + \frac{1}{4} \omega_{\mu \underline{\lambda} \underline{\sigma}} \Gamma^{\underline{\lambda} \underline{\sigma}} - i q_f A_\mu.
\end{equation}
Here $\Gamma^{\underline{\mu}}$ are the bulk $\Gamma$-matrices and $ \Gamma^{\underline{\lambda} \underline{\sigma}} =\Gamma^{[\underline{\lambda}}\Gamma^{\underline{\sigma}]}$, $e^{\mu}_{\underline{\lambda}}$ is the vielbien (we consider the square root of the diagonal metric \eqref{equ:ds_ansatz}) and $\omega$ is the spin connection. 

Note that we choose the charge of the fermion to be half of the charge of the Proca field ($q_f=1$), so that the quadratic interaction terms are allowed in \eqref{equ:fermion_action}. We also set $m_f = 0$ in our examples. We can consider 3 possible interaction terms, which would couple the bulk fermions to the vector order parameter:
\begin{align}
\label{equ:fermion_interaction_terms}
&\mbox{$\Gamma_{\mu}$-coupling:} & & \lambda_{V} B_{\mu} e^{\mu}_{\underline{m}} \bar{\Psi} \Gamma^{\underline{m}} \Psi^c + c.c., \\
\notag
&\mbox{$D_{\mu}$-coupling:} & & \lambda_{D} B_{\mu} g^{\mu \nu} \bar{\Psi} D_{\mu} \Psi^c + c.c., \\
\notag
&\mbox{$\Gamma_{\mu \nu}$-coupling:} & & i \lambda_{T} W_{\mu \nu} e^{\mu}_{\underline{m}} e^{\nu}_{\underline{n}} \bar{\Psi} \Gamma^{\underline{m} \underline{n}} \Psi^c + c.c.
\end{align}
As we discuss in more detail in Appendix \ref{app:fermion}, all these couplings are allowed on the symmetry grounds. They also would generically lead to opening the gap in the fermionic spectrum in case the order parameter condenses in a spatial direction (i.e. $B_x \neq 0$) realizing the conventional P-wave superconductor state. In this case the gap will also have two nodes in the direction along the order parameter, which corresponds to ``phase A'' P-wave superconductors. In the case of $P + i P$-condensation, discussed above, the fermion spectrum would be gapped isotropically everywhere on the Fermi surface, realizing the conventional behavior of the node-less P-wave superconductor.

\begin{figure}
\raisebox{-0.5\height}{\includegraphics[width=0.34\linewidth]{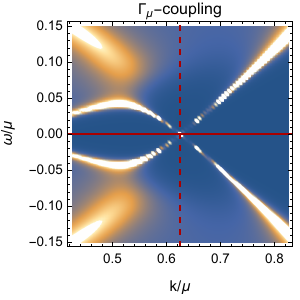}}
\raisebox{-0.5\height}{\includegraphics[width=0.31\linewidth]{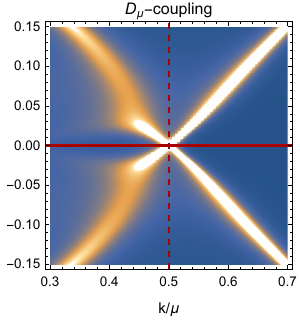}}
\raisebox{-0.5\height}{\includegraphics[width=0.31\linewidth]{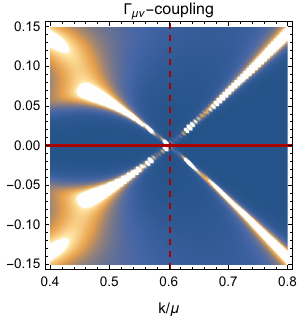}}
\centering
\raisebox{-0.5\height}{\includegraphics[width=0.34\linewidth]{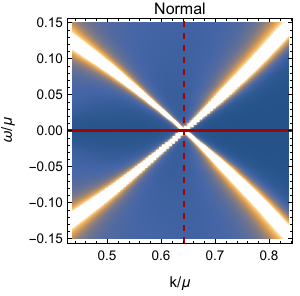}}
\raisebox{-0.5\height}{\includegraphics[width=0.31\linewidth]{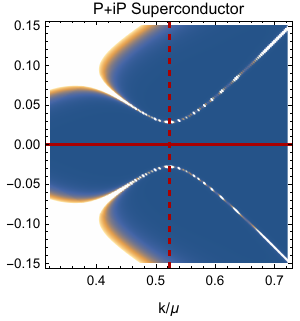}}

\caption{\label{fig:fermi_tlike} \textbf{Fermionic spectral density} in timelike superconducting state (top row) and in reference normal state (bottom row, left) and $P+i P$ superconducting state (bottom row, right). All the states are considered at $T/\mu=0.005$ and model parameters \eqref{equ:parameters}. The bulk fermion parameters are $m_f = 0, \, q_f= 1$. In the top row the different couplings \eqref{equ:fermion_interaction_terms} are switched on one-at-a-time: $\lambda_V = 1, \lambda_D = 1, \lambda_T = 0.1$, correspondingly. In all cases there is no gap in the spectrum, unlike the more conventional $P + i P$-state (here we use $\lambda_V = 0.2$), where the gap is evident. The Fermi momentum is different in all cases and is shown by the dashed red gridline.}
\end{figure}

Since our model supports both spatial and time-like condensation of the order parameter, we can study the effect of the timelike superconductivity on the fermionic spectrum, using \textit{exactly the same} interaction terms \eqref{equ:fermion_interaction_terms}, which lead to a conventional phenomenology for the spatial P-wave state. The results of our investigation are shown on Fig.\,\ref{fig:fermi_tlike}. We take the timelike SC state at $T/\mu = 0.005$ at the model parameters \eqref{equ:parameters}, the same as for the transport studies in the previous section. We consider the spectral function of the fermionic operator, turning on the interactions \eqref{equ:fermion_interaction_terms} one-by-one, what allows us to study their effect separately. The obvious crucial feature of all the results, shown on the top row of Fig.\ref{fig:fermi_tlike}, is that for all the considered couplings \textit{the gap in the fermionic spectrum never opens in the timelike superconducting state}. This behavior, while surprising on a first glance, should actually be expected, since the timelike order parameter $B_t$ is time-reversal odd and therefore it can not affect the spectrum at the symmetric point $\omega = 0$, where the usual superconducting gap would open \cite{balatsky1994properties}. For reference, we show how the conventional gap looks like in the $P + i P$ - state of our model, see the right panel in the bottom row of Fig.\,\ref{fig:fermi_tlike}.
Noteworthy the absence of the gap in the fermionic spectrum is perfectly consistent with our results for transport in the previous section: in the timelike SC ordered state there remains a finite density of electrons at the Fermi surface, which contribute to the metallic Drude peak in the AC conductivity.    
At finite frequencies, however, we clearly see on the plots on the first row of Fig.\,\ref{fig:fermi_tlike} that the various couplings affect the dispersion and the spectral gaps are opening in some cases, depending on the type of the coupling \eqref{equ:fermion_interaction_terms}. For reference we show the spectral density in the normal metallic state ($B_t = B_z = 0$) on the left panel of the bottom row of Fig.\,\ref{fig:fermi_tlike}.

We can analyze the fermionic response in more detail by fitting the data shown on Fig.\,\ref{fig:fermi_tlike} with Lorentzians at every fixed frequency slice, i.e. fitting momentum distribution curves (MDC). Note that the fit is not perfect, since in the holographic model the finite background spectral density is present even outside of the dispersion curves. The latter doesn't obstruct fitting the sharp and high peaks, which mark the fermionic dispersion. For each frequency level we use the fitting ansatz with two peaks ("inner" and "outer", related to the Fermi surface)
\begin{equation}
\label{equ:fermion_fitting}
\rho (\omega, k) = \frac{A_{in}(\omega) \Sigma_{in}(\omega)^2}{(k - k_f - v_{in}(\omega) \omega)^2 + \Sigma_{in}(\omega)^2 } +  \frac{A_{out}(\omega) \Sigma_{out}(\omega)^2}{(k - k_f - v_{out}(\omega) \omega)^2 + \Sigma_{out}(\omega)^2 }.
\end{equation}

\begin{figure}
\raisebox{-0.5\height}{\includegraphics[width=0.32\linewidth]{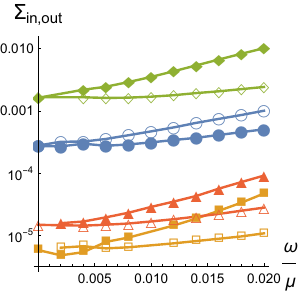}}
\raisebox{-0.5\height}{\includegraphics[width=0.32\linewidth]{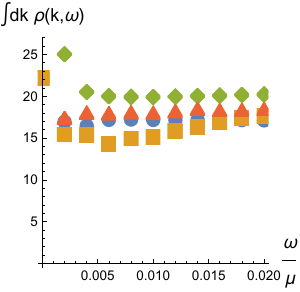}}
\raisebox{-0.5\height}{\includegraphics[width=0.32\linewidth]{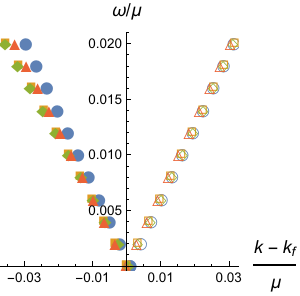}}
\centering
\raisebox{-0.5\height}{\includegraphics[width=0.8\linewidth]{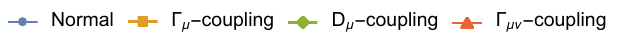}}
\caption{\label{fig:fermi_fits} 
\textbf{Fermionic spectral density fitting parameters} \eqref{equ:fermion_fitting}. The width of the peak $\Sigma$ characterizing the self energy -- left panel (note logarithmic scale), the momentum integrated spectral density \eqref{equ:integrated_spectrum} -- middle panel, and the dispersion of the peak maxima at small frequency -- right panel. The filled markers denote the inner peaks and the empty markers denote the outer peak related to the Fermi surface. Similarly to Fig.\,\ref{fig:fermi_tlike}, the different couplings in timelike superconducting state, as well as the normal state are analyzed. }
\end{figure}

Given this parametrization we can extract some features of the fermion spectral function, shown on Fig.\,\ref{fig:fermi_fits}. Firstly, as one can already notice on the density plots of Fig.\,\ref{fig:fermi_tlike}, we observe that the width of the peaks $\Sigma$ behaves quite differently at small frequencies, depending on the type of the coupling. On the left panel of Fig.\ref{fig:fermi_fits} we show the width of the peaks for different couplings (filled markers -- inner peak, empty markers -- outer peak). Note the logarithmic scale on the vertical axis. Clearly, the width for different couplings changes by orders of magnitude\footnote{The extremely narrow peaks in case of the $\Gamma_{\mu}$-coupling affect the quality of the fit.}, however the slopes of the frequency dependence in all the fits in timelike state (yellow, green and red markers) are similar, while different from the one of the normal state (blue markers). The dependence of the width of the peak on frequency close to the Fermi surface is primarily controlled by the near-horizon geometry of the background \cite{Faulkner:2009wj,Zaanen:2015oix}, rather then the type of the bulk coupling. The backgrounds we use for timelike and normal states are different and therefore the observed behavior is consistent.

Further discussing the absence of the gap in the timelike SC state it is interesting to investigate the integrated spectral density of the fermions. Given the fits \eqref{equ:fermion_fitting} and taking into account the integration along the Fermi surface we evaluate it as
\begin{equation}
\label{equ:integrated_spectrum}
\int d^2\vec{k} \, \rho(\vec{k},\omega) = \sum \limits_{a = \{in, out\}} 2 \pi \left(k_f + v_{a}(\omega) \, \omega \right) \cdot  \pi A_{a}(\omega) \Sigma_{a}(\omega).
\end{equation}
The results for the momentum integrated spectral density (the middle panel of Fig.\,\ref{fig:fermi_fits}) show that despite the narrowing of the peaks at small $\omega$, the spectral density contained in them is roughly constant. So we conclude that, not only there is no gap in the timelike SC state, but it doesn't even display a pseudogap or nodal behavior with suppression of the spectral weight towards the Fermi surface. 

Finally, we inspect the linear dispersion near the Fermi surface, which we observe at small frequencies ($\omega/\mu < 0.02$). Note (see the right panel of Fig.\,\ref{fig:fermi_fits}, empty markers) that the Fermi velocity for the outer peaks for all the different couplings in timelike state is the same as for the normal state. However, for the inner peaks (filled markers on Fig.\,\ref{fig:fermi_fits}) the Fermi velocities for all the couplings in timelike state are similar to each other, but notably different from the one in the normal state. Importantly, when evaluating the dispersion in the normal state, we do not just set all the couplings \eqref{equ:fermion_interaction_terms} to zero, but rather consider the different background solution for the normal state, which is characterized by the absence of $B_\mu$ - field and different profiles of the metric components. One concludes here, that the Fermi velocity is affected in the timelike superconductor state and it happens independently of the choice of the interaction terms in the bulk Dirac equations, and mostly due to the change of the gravitational profile, which is dictated by the condensation pattern of the order parameter itself. This effect of changing the Fermi velocity is similar to the one expected for T-odd superconductors \cite{balatsky1994properties}, but happens here in quite an indirect way.

\section{\label{sec:meissner}Meissner effect}
An important feature of any superconducting state is the Meissner effect. In conventional superconductor the magnetic field is repelled from the bulk of the material and decays exponentially with the rate set by the magnetic penetration length $l_m$. In time-reversal odd superconductors \cite{balatsky1994properties} the character of the Meissner effect is debated and one can argue that it has an opposite sign, leading to the instability of the state itself. Given that the timelike superconductor, which we explore in this work, is a representative of the T-odd class, it is interesting to check the character of the Meissner effect in our holographic model. 

The perturbative treatment of the Meissner effect in holographic model of the scalar ``s-wave'' superconductor has been discussed recently in \cite{Natsuume:2022kic}. By performing the double expansion in both the amplitude of the order parameter \cite{Herzog:2010vz} and spatial wave-vector\footnote{We thank the authors of \cite{Natsuume:2022kic} for clarifying this to us} in the linearized equations of motion and assuming the mixed boundary conditions \cite{Compere:2008us,Ecker:2021cvz,Jeong:2023las}, one can qualitatively analyze the Meissner effect in timelike superconductor given only the profiles of the homogeneous background solution, as shown on Fig.\,\ref{fig:Back_solutions}. We follow this approach, leaving the more sophisticated analysis of the nonlinear spatially inhomogeneous solutions, like  \cite{Albash:2009iq,Montull:2009fe}, to future work.

In this section we restrict ourselves to the probe case and set the background metric to that of the Schwarzchild black hole: 
\begin{equation}
T(z)=W(z)=U(z)=1, \qquad f(z) = 1- z^3
\end{equation}
in \eqref{equ:ds_ansatz}. In order to describe small external magnetic field $B \sim \p_x \delta A_y$, we consider the $x$-dependent perturbation of the $A_y$ component of the gauge field $\delta A_y(x,z)$.
In case when the background is spatially homogeneous, the perturbation can be represented as a spatial convolution of the boundary value with the bulk-to-boundary propagator, which in momentum space reads
\begin{equation}
\label{equ:bulk_to_boudary_prop}
\delta A_y(k,z) = \mathcal{A}_y(k) \delta \tilde{A}_y(k,z), \qquad \delta \tilde{A}_y(k,z)\Big|_{z\rar0} = 1. 
\end{equation}
Following \cite{Natsuume:2022kic} in order to enable the dynamical electromagnetism on the boundary, we impose the  Maxwell equation, which couples the boundary electromagnetic field, the expectation value of the current and the external source $\hat{J}^i$.
\begin{gather}
\label{equ:semiclassical_maxwell}
\p_i \mathcal{F}^{ij} = e^2 \la \mathcal{J}^i \ra + e^2 \hat{J}^i\\
\notag
\mathcal{F}^{ij} = \p_i \mathcal{A}^j -  \p_j \mathcal{A}^i \\
\notag
\la \mathcal{J}^i \ra = \mathcal{A}^i(k) \p_z \delta \tilde{A}^i(k,z)\Big|_{z\rar0} 
\end{gather}
Given $\delta \tilde{A}^i(p,z)$ one solves \eqref{equ:semiclassical_maxwell} and obtains the spatial profile of the boundary gauge field $\mathcal{A}^i$ in momentum representation.

At the linear order, in presence of the timelike condensate $B_t, B_z$, the perturbation $\delta \tilde{A}^y(k,z)$ couples to $y$-component of the charged vector field:
\begin{align}
\label{equ:dAy_equations_timelike}
\p_z \left(f \p_z \delta \tilde{A}_y  \right) - & k^2 \delta \tilde{A}_y + \frac{2 q_B^2}{f} \left(B_t^2 - f^2 B_z^2 \right) \delta \tilde{A}_y \\
\notag
+ & 4 q_B f B_z \, \p_z \delta B_y^R + 2 q_B \left( \p_z (f B_z) - \frac{q_B}{f} A_t B_t \right) \delta B_y^R = 0
\\
\notag
\p_z \left(f \p_z \delta \tilde{B}_y^R  \right) - & k^2 \delta \tilde{B}^R_y - \frac{q_B^2 \phi^2}{z^2} \tilde{B}^R_y + \frac{q_B^2}{f} A_t^2 \delta \tilde{B}_y^R - 2 q_B^2 B_z^2 \delta \tilde{B}_y^R \\
\notag
- & 2 f q_B B_z \p_z \delta \tilde{A}_y - q_B \left( \p_z (f B_z) + \frac{q_B}{f} A_t B_t \right)  \delta \tilde{A}_y=0
\end{align}
where $\delta B_y^R$ denotes the real part of the $\delta B_y$-component and the imaginary part decouples.

\begin{figure}
\raisebox{-0.5\height}{\includegraphics[width=0.4\linewidth]{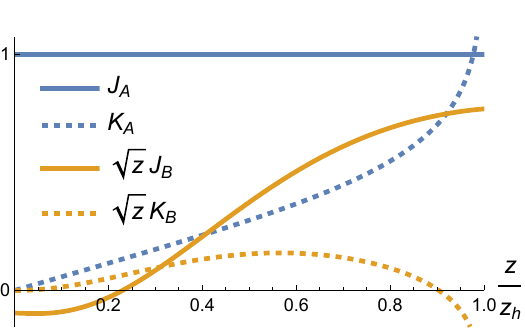}}
\raisebox{-0.5\height}{\includegraphics[width=0.49\linewidth]{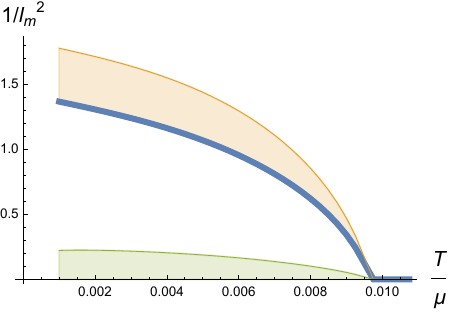}}
\caption{\label{fig:meissner_results} 
\textbf{Magnetic penetration length in timelike superconductor.} \textit{Left panel:} The pairs of linearly independent basis used to construct the Green's function in the parturbative calculation of the Meissner effect. Note that the $K_B, J_B$ profiles are rescaled for presentation purposes. $K_A, K_B$-profiles are defined as being zero at the boundary $z=0$, $J_A, J_B$ are finite at the horizon. See Appendix\,\ref{app:Green_functions}. \\
\textit{Right panel:} The inverse magnetic penetration length as a function of temperature (thick blue line). The inverse penetration length is positive and grows with the order parameter.  The shaded yellow area shows the negative contribution due to $B_t^2$-term in \eqref{equ:value_of_I}. The green shaded area shows the positive contribution of the $\mathcal{H}$-term.}
\end{figure}

We can perturbatively analyze the solution to these equations in the regime where the condensate is weak: $B_t, B_z \sim \epsilon, \epsilon \rar 0$ and focusing on the long wavelength limit $k \rar 0$. One can employ the Green's function method, which we describe in more detail in Appendix \ref{app:Green_functions} in order to obtain the solution in the form
\begin{equation}
\label{equ:perturbative_expansion_Meissner}
\delta \tilde{A}^y(p,z) = \delta \tilde{A}^{y(0,0)}(z) + k^2 \delta \tilde{A}^{y(0,2)}(z) + \epsilon^2 \delta \tilde{A}^{y(2,0)}(z).
\end{equation}
Plugging this into the definition of the current expectation value and further in \eqref{equ:semiclassical_maxwell} we obtain the equation on $\mathcal{A}(k)$ in the following form
\begin{equation}
\label{equ:maxwell_with_I}
k^2 \mathcal{A}_y(k) = - e^2 \mathcal{A}_y(k) \left(k^2 + 2 I \right) +  e^2 \hat{J}^i,
\end{equation}
where $I$ is the integral
\begin{align}
\label{equ:value_of_I}
I &= \int \limits_{0}^{1} dz' \frac{q_B^2}{f(z')} \left(- B_t(z')^2 + f(z')^2 B_z(z')^2 \right) \\
\notag
& + \int \limits_{0}^{1} dz' \mathcal{H}(z') \left[\frac{J_B(z')}{C_B} \int \limits_{0}^{z'} ds K_B (s) \mathcal{H}(s) +  \frac{K_B(z')}{C_B} \int \limits_{z'}^{1} ds J_B (s) \mathcal{H}(s) \right] \\
\notag
\mathcal{H}(s) &\equiv  q_B \left( \p_{s} (f(s) B_z(s)) + \frac{q_B}{f(s)} A_t(s) B_t(s) \right)
\end{align}
Here $K_{A,B}(z)$ and $J_{A,B}(z)$ are the linearly independent solutions to the corresponding equations of motion at leading order (see Appendix \ref{app:Green_functions}) and $C_B = f (K_B J'_B - J_B K_B')$ is the constant in the corresponding Wronskian. We show their profiles on the left panel of Fig.\,\ref{fig:meissner_results}. The value of $I$ is controlled by the profiles of the background fields responsible for the timelike order parameter: $B_t$ and $B_z$.

Note that in presence of the external source, the solution to \eqref{equ:maxwell_with_I} in position space is
\begin{equation}
\mathcal{A}^y(x) \sim \hat{J}^y e^{- x/l_m}, \qquad l_m^2 = \frac{1+e^2}{2 e^2 I }.
\end{equation}
Here $l_m$ is the magnetic penetration length, which is inversely proportional to the integral $I$. It is crucial that $I$ in \eqref{equ:maxwell_with_I} contributes at the super-leading order in $k^2$ and therefore it changes the qualitative behavior of $\mathcal{A}^y$ at large distances already at the perturative level in the amplitude of the order parameter $\epsilon$. (In absence of $I$ term the $\mathcal{A}^y$ would not decay at all.)

As it has been shown in \cite{Natsuume:2022kic}, and as we review in the Appendix \ref{app:Green_functions}, in case of the scalar holographic superconductor the corresponding integral is positive definite and therefore the Meissner effect in scalar model is always physically consistent. In our case of timelike order parameter the situation is more subtle, since the sign of the integral \eqref{equ:value_of_I} depends on the competition between the terms proportional to $B_t^2$ and $B_z^2$. Note that $B_z$-contribution here is crucial, since otherwise the integral would be negative definite. Evaluating the integral in the timelike superconducting state at various temperatures, we observe (see the right panel of Fig.\,\ref{fig:meissner_results}) that the inverse penetration length is positive and grows as the order parameter increases. This should be expected for the physically sensible state. We also show that the $B_z^2$ term in the integral is dominating and, while $B_t^2$ gives a comparable negative contribution, the overall sum remains positive.   

\section{\label{sec:conclusion}Conclusion}

In this work we introduced the consistent holographic model which exhibits condensation of the timlike vector superconducting order parameter and studied the phenomenological features of this novel state of quantum matter. 
We explored both thermodynamical and dynamical stability of this state and showed that it is at least metastable. We showed that the state exhibits a positive Meissner effect in all the studied examples. Unlike the scalar superconductor, however, the sign of the Meissner effect is not generically protected.

Our study of the AC thermoelectric transport reveals an interesting feature of this state, which is similar to a gapless superconductor: the timelike SC state supports the supercurrent, but simultaneously exhibits the metallic Drude peak in the AC spectrum. The gap in the AC spectrum, characteristic to the conventional superconductors is absent, but instead the spectral weight transfer to the higher frequencies is revealed by the kink-like feature in the spectrum.

The absence of the gap is also evident in the fermionic spectral function. We show that irrespective of the type of the bulk fermionic coupling, the gap is never present and the integrated spectral density at the Fermi surface is finite. We demonstrate that the Fermi velocity is affected by the presence of the timelike condensate. It happens in an indirect way due to the backreaction on the geometry and the effect is generic, since it doesn't depend on the type of the bulk fermionic coupling.  

The model which we construct derives from the $SU(2)$-gauge symmetry, which is explicitly broken by the boundary value of the adjoint Higgs field. Correspondingly the boundary theory, which we describe, contains the structure of $SU(2)$ global symmetry, which is explicitly broken by the source of the scalar operator. It would be interesting to try to interpret this broken $SU(2)$ from the physical perspective.  

As we observe for the chosen model parameters \eqref{equ:parameters}, the timlike SC state is only thermodynamically metastable, while $P + i P$ SC state has a higher critical temperature and is thermodynamically preferred. The interplay between the different states, however, depends on the model parameters: the charge of the vector field $q_B$ and the boundary value of the scalar field $\phi$, controlling the mass of the vector field.
It will be interesting to further explore the parameter space and figure our the regimes where either timelike or P-wave state is thermodynamically preferred. We also note that in present case we couldn't find a stable anisotropic P-wave solution, it might be that such solution would be promoted by the periodic crystal lattice, instead of the isotropic Q-lattice background. The study of this more realistic setup will be performed elsewhere. 

Given that in our model the mass of the vector field is controlled by the boundary source, it is particularly suitable for studying various heterostructures and spatially inhomogeneous setups, where the various phases of the superconductor are realized in different spacial regions, controlled by the source of $\phi$. The elliptic equations of motion, which we use for this model, can be straightforwardly generalized to several spatial directions and implemented on the corresponding numerical grid. This generalization will allow for comparison to the experimental studies of the transport across heterostructures, where the T-odd state is sandwiched between two conventional superconductors \cite{tanaka2007anomalous}. In the same inhomogeneous approach one would also be able to study the nonlinear Meissner effect, when considering the droplet of an SC state \cite{Albash:2009iq}.

It would be interesting to further explore the features of the fermionic spectrum at the Fermi surface, in particular the self energy and the Fermi velocity. This can be done by constructing the zero temperature geometries in our model and studying their near IR behavior along the lines of \cite{Faulkner:2009wj,Rodgers:2022uvs}. The structure of the equations of motion in our model allows for such a study.

Overall, in this work we demonstrate that holography, and in particular our model, provide a useful framework for studying unconventional features of timelike, and more generally T-odd, superconducting states. 

\acknowledgments
We thank Alexander Balatsky to bringing this class of problems to our attention and acknowledge the participation of Cristiana Pantilidou at the preliminary discussions on this project several years ago. A.K. is grateful to Koenraad Schalm, Kostya Zarembo, Elias Kiritsis, Tomas Andrade, Annica Black-Shaffer and Ronnie Rodgers for useful comments. A.K. particularly benefited from the advice of Fawad Hassan regarding the massive vector fields and the one of Sergey Moroz regarding the $P+ i P$-state. A.K. also acknowledges the collaboration with Alexander Pershoguba and Christopher Triola on the related project which has been inspirational, but has never been completed. 

The work of A.K. is supported by VR Starting Grant 2018-04542 of Swedish Research Council.

A.K. appreciates the opportunity to deliver the preliminary results of this study at ``Zaanen's Fest'' (
Stripes, Planckian Dissipation and Quantum supremacy, 17 \& 18 July 2023, Leiden) and acknowledges the useful feedback provided by the participants. We also benefited a lot from the feedback obtained during the local Nordita seminars hosted by HEP and CMT groups.

U.E. acknowledges the hospitality of Nordita within Nordita visitng PhD fellow program.
The visit of U.E. to Nordita is partially supported by Beca de Movilidad Internacional para alumnos de especializaci\/on, maestr\/ia y doctorado, Universidad Aut\/onoma Metropolitana.

The computations were enabled by resources provided by the National Academic Infrastructure for Supercomputing in Sweden (NAISS) at PDC Center for High Performance Computing partially funded by the Swedish Research Council through grant agreement no. 2022-06725.

% \begin{equation}
% \label{eq:Dirac_equation}
% \Gamma^{\underline{a}} e_{\underline{a}}^\mu \left( \p_\mu + \frac{1}{4} \omega_{\m\underline{bc}}\Gamma^{\underline{bc}} - i q A_\mu \right) \Xi - m \Xi =0, 
% \end{equation}

\appendix

\section{\label{app:charged_Proca}Equations of motion for the charged vector field}
In this section we will discuss the possible choices for the action in the model. Our task is to find an action for the charged vector field, which would be consistent and describe the correct number of degrees of freedom in presence of external fields (we partially follow the logic of the treatment in Appendix A of \cite{Benini:2010pr}) and, on the other side, would support the condensation of the timelike component of the vector field. 

\subsection*{The neutral Proca field}
We start by looking into the conventional Proca action which describes a neutral massive vector $P_{\mu}$.
\begin{gather}
\eqref{equ:neutral_Proca_action}
S_{Proca} = \int d^4 x \sqrt{-g} \left( \frac{1}{2} H_{\mu \nu} H^{\mu \nu} - m_P^2 P_{\mu} P^{\mu} \right) \\
H_{\mu \nu} = \nabla_{\mu} P_{\nu} - \nabla_{\nu} P_{\mu}.
\end{gather}
The equations of motion read
\begin{equation}
\label{equ:neutral_Proca_EOMS}
\nabla_{\mu} H^{\mu \nu} - m_P^2 P^{\nu} = 0.
\end{equation}
There are four equations on the four components of the vector field. However, the massive spin 1 particle has 3 physical degrees of freedom, therefore we need to find two constraints to remove the unnecessary propagating mode. The $z$-component of \eqref{equ:neutral_Proca_EOMS} does not contain the second order $z-$ derivative and therefore provides one constraint. The other constraint is obtained when taking the divergence of \eqref{equ:neutral_Proca_EOMS}:
\begin{equation}
\label{equ:dP_constraint}
\nabla_{\nu} \nabla_{\mu} H^{\mu \nu} = m_P^2 \nabla_{\nu} P^{\nu} = 0.
\end{equation}
The last equality is due to the symmetry of double covariant derivative of the antisymmetric tensor $H^{\mu \nu}$. The two constraints remove the unphysical degree of freedom and one is left with the correct dynamics: 3 propagating modes, characterized by the 6 integration constants.

In practice, when solving the boundary value problem one can impose the constraint on the boundary and add the constraint conservation equation (the divergence \eqref{equ:dP_constraint}) to the equations of motion \eqref{equ:neutral_Proca_EOMS}. This leads to a set of four regular second order equations
\begin{equation}
\nabla_{\mu} \nabla^{\mu} P_{\nu} - m_P^2 P^{\nu} = 0.
\end{equation}
These equations can be solved subject to the appropriate boundary conditions, respecting the constraint \eqref{equ:dP_constraint} which yields a solution to \eqref{equ:neutral_Proca_EOMS}.

\subsection*{\label{app:charged_Proca}\label{sec:Proca_EOMs}The charged Proca model}
In \cite{Cai:2013aca, Arias:2016nww} a generalization of the neutral Proca model \eqref{equ:neutral_Proca_action} has been suggested, promoting the real vector $P_\mu$ to a complex one $B_\mu$, which is charged under the U(1) gauge field $A_{\mu}$. The action is obtained from \eqref{equ:neutral_Proca_action} by replacing all the derivatives with the gauge covariant ones.
\begin{gather}
\label{equ:neutral_Proca_action}
S_{\mathrm{charged\, Proca}} = \int d^4 x \sqrt{-g} \bigg( - \frac{1}{4} F_{\mu \nu} F^{\mu \nu} - \frac{1}{2} W_{\mu \nu} \bar{W}^{\mu \nu} - m_B^2 B_{\mu} \bar{B}^{\mu} \bigg) \\
W_{\mu \nu} = D_{\mu} B_{\nu} - D_{\nu} B_{\mu}, \qquad D_{\mu} B_{\nu} = \nabla_{\mu} B_{\nu} - i q A_{\mu} B_{\nu}
\end{gather}
The equations of motion in this theory are 
\begin{equation}
\label{equ:Proca_eom}
D_{\alpha} {W}^{\alpha \beta} = m_B^2 {B}^\beta, \qquad \mathrm{and \ c.c.}
\end{equation}
Similarly to the neutral case, the $z$ components of the equations do not contain derivatives and therefore provide the constraints, which remove one integration constant. The divergence of the equations of motion now contains the extra terms involving the derivatives of the gauge field, which come from the commutation of the covariant derivatives. 
\begin{equation}
\label{equ:DB_constraint}
m_B^2 D_{\nu} B^{\nu} = D_{\nu} D_{\mu} W^{\mu \nu} \sim q_B \{\p A \}.
\end{equation}
When considering the constraint conservation equation we can take a derivative of \eqref{equ:DB_constraint} and add it to \eqref{equ:Proca_eom} in order to obtain
\begin{equation}
\label{equ:enhanced_Proca}
D_{\alpha} {W}^{\alpha \beta} + D^{\beta} D_{\alpha} B^{\alpha}= m_B^2 {B}^\beta +  \frac{1}{m_B^2} q_B \{\p \p A B\, \}.
\end{equation}
This system of equations contains second order derivatives of both $B_t(z)$ and $B_z(z)$, which is suitable for a well defined numerical elliptic problem. However the extra complication is the presence of second order derivative terms of $A$ in the right hand side. When diagonalizing the full system of the equations of motion with respect to the second derivatives, one encounters the equations with irregular coefficients, which depend on the background fields and may contain singularities, where the coefficients in front of the highest differential operators vanish. This leads to the occasional instabilities.

\subsection*{\label{app:non_Abelian}The non-Abelian model}

Another way to construct the Lagrangian for the charged vector field is to rely of the non-Abelian $SU(2)$ gauge model discussed in \cite{Gubser:2008wv} 
\begin{equation}
\label{equ:SU2_pure}
S = \int d^4 x - \frac{1}{2} \tr \mathcal{G}_{\mu \nu}^{\dag} \mathcal{G}^{\mu \nu} 
\end{equation}
\begin{align}
\mathcal{G}_{\mu \nu} &= \p_\mu \mathcal{B}_{\nu} - \p_\nu \mathcal{B}_{\mu} - i q [\mathcal{B}_\mu, \mathcal{B}_\nu], 
&
\mathcal{B}_{\mu} &= B_{\mu} \tau^+ + \bar{B}_{\mu} \tau^- + A_{\mu} \tau^3.
\end{align}
This is evidently consistent. The extra constraints, needed to remove the unphysical degrees of freedom come from the gauge symmetry and the conservation of constraints follows from the equations of motion themselves. This model, however, does not allow to include the mass of the vector field. Because of this extra rigidity, one can show that the nontrivial solution for $B_t$-field is excluded by the gauge constraint. Indeed, in case of the radial dependence of the fields, in our metric ansatz \eqref{equ:ds_ansatz} and assuming the gauge $B_z = 0$ one reads from the equations of motion:
\begin{gather}
\p_z^2 B_t (z) + \p_z B_t \left(-\frac{T'}{2 T}  -\frac{U'}{2 U} + \frac{W'}{2 W}\right) = 0, \\
B_t A_t' - A_t B_t' = 0
\end{gather}
Where the latter is the gauge constraint. These equations have only a trivial solution for $B_t$. 

\subsection*{\label{app:adjoin_Higgs}The non-Abelian model with adjoint Higgs}

Once we add an extra scalar field in adjoint representation, the rigidity of the model \eqref{equ:SU2_pure} is relaxed and the $B_t$-field now couples to the scalar
\begin{gather}
\p_z^2 B_t (z) + \p_z B_t \left(-\frac{T'}{2 T}  -\frac{U'}{2 U} + \frac{W'}{2 W}\right) + \frac{q_B^2 A_t \phi \Phi U}{f z^2} - \frac{q_B^2 B_t \phi^2 U}{f z^2} = 0,
\end{gather}
and acquires the necessary nontrivial dynamics.

On the other hand, unlike the charged Proca case, the $SU(2)$ model with adjoint Higgs field retains the gauge symmetry at the level of equations of motion and thereofore has the appropriate number of constraints and dynamical degrees of freedom of the charged massive vector boson (this is a direct analogue of W-boson in Standard model). The constraint conservation equation doesn't conflict with the remaining equations ans the boundary value problem can be consistently formulated.

\section{\label{app:fermi_couplings}Fermionic coupling terms}

%recent discussion on this \cite{Ghorai:2023wpu}

In this Appendix we study different options for the coupling between the probe fermion and the background charged Proca field. We will look in particular into the interaction terms which generically lead to opening of the superconducting spectral gap in the conventional $P$-wave state. In this analysis we follow \cite{Faulkner:2009am,Benini:2010qc}. To begin with, let us assume the following representation for the bulk $\Gamma$-matrices \cite{Benini:2010qc}:
\begin{equation}
\Gamma^{\underline{t}} = \begin{pmatrix} - i \sigma_2 & 0 \\ 0 & i \sigma_2 \end{pmatrix}, \quad
\Gamma^{\underline{x}} = \begin{pmatrix} \sigma_1 & 0 \\ 0 & \sigma_1 \end{pmatrix}, \quad
\Gamma^{\underline{y}} = \begin{pmatrix} 0 & - i \sigma_2 \\ i \sigma_2 & 0 \end{pmatrix}, \quad
\Gamma^{\underline{z}} = \begin{pmatrix} \sigma_3 & 0 \\ 0 & \sigma_3 \end{pmatrix}. 
\end{equation}
Defining $\Gamma^5 \equiv i \Gamma^{\underline{t}}  \Gamma^{\underline{x}}  \Gamma^{\underline{y}}  \Gamma^{\underline{z}}$ and charge conjugation as $C \Gamma^{\underline{\mu}} C^{-1} = - {\Gamma^{\underline{\mu}}}^{T}$, we have
\begin{equation}
 \Gamma^{5} = \begin{pmatrix} 0 & \sigma_2 \\ \sigma_2 & 0 \end{pmatrix}, \qquad C = \Gamma^{\underline{t}} = \begin{pmatrix} - i \sigma_2 & 0 \\ 0 & i \sigma_2 \end{pmatrix}.
\end{equation}
The superconducting interaction terms in the Dirac equation mix the positive and negative frequencies, therefore for the rescaled bulk spinor field ($\psi = (-g \cdot g^{zz})^{1/4} \Psi$) we consider the ansatz
\begin{equation}
\psi = e^{-i \omega t + i \vec{k} \cdot \vec{x}} \zeta + e^{i \omega t - i \vec{k} \cdot \vec{x}} \tilde{\zeta}.
\end{equation}
Moreover, we split $\zeta$ and $\tilde{\zeta}$ into two two-component parts $\zeta = (\zeta_1, \zeta_2)^T$. 

In absence of the interaction terms and in case $k_y = 0$, the Dirac equations for $\zeta_1$ and (complex conjugate) $\tilde{\zeta}_2^*$ are particularly simple
\begin{align}
\left[\p_z  + \sqrt{\frac{U}{f^2 T}} (\omega + q_f A_t) i \sigma_1 - \sqrt{\frac{U}{f W}} k_x \sigma_2 - \sqrt{\frac{U}{f z^2}} m_f \sigma_3 \right] & \zeta_1 \equiv \mathcal{D}_\omega \zeta_1 = 0,  \\
\left[\p_z  + \sqrt{\frac{U}{f^2 T}} (-\omega + q_f A_t) i \sigma_1 - \sqrt{\frac{U}{f W}} k_x \sigma_2 - \sqrt{\frac{U}{f z^2}} m_f \sigma_3 \right] & \tilde{\zeta}_2^* \equiv \mathcal{D}_{-\omega} \tilde{\zeta}_2^* = 0. 
\end{align}
It follows that the dispersion relation of $\tilde{\zeta}_2^*$ is a reflection of the dispersion of $\zeta_1$ about the $k_x$ axis, and they generically intersect at $\omega=0$. Therefore the coupling term which would mix $\zeta_1$ and $\tilde{\zeta}_2^*$ would lead generically to opening the spectral gap around $\omega =0$ due to eigenvalue repulsion, which is the common manifestation of superconductivity in the fermionic spectrum. 

\subsection*{Scalar superconductor}

For example, in \cite{Faulkner:2009am} the fermionic spectrum of $s$-wave holographic superconductor has been studied. The order parameter in this case is controlled by the charged scalar field $\Phi$ (with twice the charge of the fermion) and the interaction term with the following structure is allowed
\begin{equation}
\label{equ:s-coupling}
\mathit{L}^S_{int} = \Phi \bar{\Psi}(\eta + \eta_5 \Gamma^5) \Psi^c + c.c., 
\end{equation}
where the charge conjugated spinor is $\Psi^c \equiv C \Gamma^{\underline{t}} \Psi^*$, Dirac conjugation is $\bar{\Psi} \equiv (\Psi^*)^T \Gamma^{\underline{t}}$ and $\{*,T\}$ are complex conjugation and transposition operators, respectively. Given these definitions, the coupling \eqref{equ:s-coupling} contributes to the equation of motion for $\Psi$, and in particular $\zeta_1$ as
\begin{equation}
\mathcal{D}_\omega \zeta_1 - 2 \eta_5 \Phi \sqrt{\frac{U}{z^2 f}} \sigma_1 \tilde{\zeta}_2^* + 2 \eta \Phi \sqrt{\frac{U}{z^2 f}} i \sigma_3 \tilde{\zeta}_1^* = 0.
\end{equation}
We see that it is $\eta_5$-term, which couples generically $\zeta_1$ and $\tilde{\zeta}_2^*$, and therefore leads to opening of the superconducting gap at $\omega = 0$. In contrast, $\eta$-term couples the modes with different dispersions, which may intersect at arbitrary points away from the Fermi surface. Therefore it can lead to opening the gap at $\omega \neq 0$, which will not affect crucially the low energy spectrum. For these reasons only the effect of $\eta_5$ coupling has been considered in \cite{Faulkner:2009am} (see \cite{Benini:2010qc,Benini:2010pr} for the similar analysis in case of D-wave order).

\subsection*{Vector superconductor: $\Gamma_{\mu}$-coupling}

Now we turn to analysis of the coupling to P-wave order parameter, which we are after in this work. Given, again, that the charge of the order parameter field is twice the charge of the bulk fermion, the following interaction term is allowed \eqref{equ:p-interaction_temporal}
\begin{equation}
\label{equ:p-coupling}
\mathit{L}^V_{int} = B_\mu e^{\mu}_{\underline{m}} \bar{\Psi} \Gamma^{\underline{m}} (\lambda_{V} + \lambda_{V5} \Gamma^5) \Psi^c + c.c. 
\end{equation}
In case of the vector order parameter, the mixing of fermion modes depends on the direction in which the vector condenses. Keeping momentum of the fermions in $x$-direction, we see that the mixing introduced by $B_x$ and $B_y$ condensates reads
\begin{align}
\label{equ:p-interaction_spacial}
\mathcal{D}_\omega \zeta_1 +  2 \sqrt{\frac{U}{f W}} (- \lambda_{V5} B_x \mathbf{1}_2 + \lambda_V B_y i \sigma_1) &\tilde{\zeta}_2^* \\
\notag
+  2 \sqrt{\frac{U}{f W}} (\lambda_{V5} B_y  \sigma_3 + \lambda_V B_x \sigma_2) & \tilde{\zeta}_1^* = 0.
\end{align}

Finally, we can look at the couplings mediated by the temporal component of $B_t$, which condenses in timelike SC state. Following from the interaction term \eqref{equ:p-coupling}, the corresponding mixing reads
\begin{align}
\label{equ:p-interaction_temporal}
\mathcal{D}_\omega \zeta_1 - 2 \lambda_{V5} \left(B_z \sigma_2 - B_t \sqrt{\frac{U}{f^2 T}} \sigma_3 \right) &\tilde{\zeta}_2^* \\
\notag
+  2 \lambda_V \left(B_z \mathbf{1}_2 + B_t \sqrt{\frac{U}{f^2 T}} i \sigma_1 \right) & \tilde{\zeta}_1^* = 0.
\end{align}

Note that $\lambda_{V5}$ term in \eqref{equ:p-coupling} couples vector $B_{\mu}$ to the axial-vector current $\bar{\Psi} \Gamma^{\mu} \Gamma^5 \Psi^c$. Therefore it should be excluded by the requirement of parity symmetry of the Lagrangian. Inspecting the interaction terms in \eqref{equ:p-interaction_spacial} we see that at $\lambda_{V5} = 0$ the $p$-wave gap is generically formed in case when spatial component of the vector order parameter condenses. In particular, the gap opens in all directions except the two ``nodes'', where momentum of the electrons is parallel to the order parameter. Remarkably, in case of timelike order parameter at $\lambda_{V5}=0$ the gap does not generically open in any direction according to \eqref{equ:p-interaction_temporal}. This is a characteristic feature of the time reversal odd superconducting state, where the $U(1)$ symmetry is broken, but Fermi surface remains ungapped. 

\subsection*{Vector superconductor: $D_{\mu}$-coupling}

Similarly we can analyze a derivative coupling term in the bulk fermion Lagrangian \eqref{equ:p-interaction_temporal}
\begin{equation}
\label{equ:dp-coupling}
\mathit{L}^{D}_{int} = B^\mu \bar{\Psi} (\lambda_{D} + \lambda_{D5} \Gamma^5)D_{\mu} \Psi^c + c.c., 
\end{equation}
with covariant derivative defined as in \eqref{equ:Proca_part}. The equations of motion acquire the following contributions due to this term:
\begin{align}
\label{equ:dp-interaction_spacial}
\mathcal{D}_\omega \zeta_1 +  \left[ \lambda_{D} B_y \frac{z \p_z \log(W/z^2)}{2 \sqrt{W}} \sigma_2 - \lambda_{D5} B_x \frac{z \p_z \log(W/z^2)}{2 \sqrt{W}} \sigma_3 - \lambda_{D5} B_x \sqrt{z^2 \frac{U}{W^2 f}} 2 i k_x \sigma_1 \right] &\tilde{\zeta}_2^* \\
\notag
+  \left[ \lambda_{D5} B_y \frac{z \p_z \log(W/z^2)}{2 \sqrt{W}} \mathbf{1}_2 + \lambda_{D} B_x \frac{z \p_z \log(W/z^2)}{2 \sqrt{W}} i \sigma_1 + \lambda_{D} B_x \sqrt{\frac{z^2 U}{W^2 f}} 2 k_x \sigma_3 \right] & \tilde{\zeta}_1^* = 0
\end{align}
in case the spatial components condense, and
\begin{align}
\label{equ:dp-interaction_temporal}
\mathcal{D}_\omega \zeta_1 +  \lambda_{D5} \bigg[ B_t \frac{z^2}{f T}\left(-2 (\omega - q_f A_t) \sqrt{\frac{U}{z^2 f}} i \sigma_1 + \sqrt{\frac{f T}{z^2}} \frac{1}{2} \p_z \mathrm{log} \left(\frac{f T}{z^2}\right) \mathbf{1}_2 \right)& \\
\notag
+ B_z \sqrt{\frac{z^2 f}{U}} \left(\frac{1}{2} \p_z \mathrm{log}\left(\frac{f T W}{z^6}\right) i \sigma_1 - 2 i \sigma_1 \p_z \right)& \bigg] \tilde{\zeta}_2^* \\
\notag
+  \lambda_{D} \bigg[ B_t  \frac{z^2}{f T}\left(2 (\omega - q_f A_t) \sqrt{\frac{U}{z^2 f}} \sigma_3 + \sqrt{\frac{f T}{z^2}} \frac{1}{2} \p_z \mathrm{log} \left(\frac{f T}{z^2}\right)  \sigma_2 \right)& \\ 
\notag
+ B_z \sqrt{\frac{z^2 f}{U}} \left(- \frac{1}{2} \p_z \mathrm{log}\left(\frac{f T W}{z^6}\right) \sigma_3 + 2 \sigma_3 \p_z \right)& \bigg] \tilde{\zeta}_1^* = 0
\end{align}
in case the temporal ones do. Similarly to the vector interaction term \eqref{equ:p-coupling}, in order to preserve parity in the Lagrangian one has to set $\lambda_{D5} = 0$. In this case the p-wave gap with two nodal points is induced by the spacial components of the order parameter in \eqref{equ:dp-interaction_spacial} and no gap arises generically due to the temporal component in \eqref{equ:dp-interaction_temporal}. In the main text we do not include the derivative interaction term \eqref{equ:dp-coupling}, since it is technically more complicated, but doesn't seem to introduce extra interesting physical effects.

\subsection*{Vector superconductor: $\Gamma_{\mu \nu}$-coupling}
Finally, we consider the ``tensor'' coupling from \eqref{equ:fermion_interaction_terms}, which has the structure
\begin{equation}
\label{equ:t-coupling}
\mathit{L}^{T}_{int} = W_{\mu \nu} e^{\mu}_{\underline{m}} e^{\nu}_{\underline{n}} \bar{\Psi} \left(i \lambda_{T} \Gamma^{\underline{m} \underline{n}} + \lambda_{T5} \Gamma^5 \Gamma^{\underline{m} \underline{n}} \right)\Psi^c 
\end{equation}

In the equations of motion this coupling introduces the following terms in spacial condensation case

\begin{align}
\label{equ:t-interaction_spacial}
\mathcal{D}_\omega \zeta_1 +  \left[ \lambda_{T} \frac{4 z}{\sqrt{W}} \left(B_y' i \sigma_2   - \frac{\sqrt{U}}{f \sqrt{T}} i q_B A_t  B_y \sigma_3 \right) - \lambda_{T5}  \frac{4 z}{\sqrt{W}} \left(B_x'  \sigma_3  - \frac{\sqrt{U}}{f \sqrt{T}} i q_B A_t  B_x i \sigma_2 \right) \right] &\tilde{\zeta}_2^* \\
\notag
+  \left[ \lambda_{T} \frac{4 z}{\sqrt{W}} \left( B_x'  \sigma_1  + \frac{\sqrt{U}}{f \sqrt{T}} i q_B A_t  B_x \mathbf{1} \right) + \lambda_{T5}  \frac{4 z}{\sqrt{W}} \left( B_y'  i \sigma_2  - \frac{\sqrt{U}}{f \sqrt{T}} i q_B A_t  B_y \sigma_1 \right) \right] & \tilde{\zeta}_1^* = 0
\end{align}

And in case of the timelike condensation one obtains

\begin{align}
\label{equ:t-interaction_timelike}
\mathcal{D}_\omega \zeta_1 +  \left[  \lambda_{T5}  \frac{4 z}{\sqrt{f T}} \left(B_t' - i q_B A_t  B_z \right) \mathbf{1} \right] &\tilde{\zeta}_2^* \\
\notag
+  \left[\lambda_{T} \frac{4 z}{\sqrt{f T}} \left(B_t' - i q_B A_t B_z\right) i \sigma_2 \right] & \tilde{\zeta}_1^* = 0
\end{align}

Similarly to the other cases above, one has to drop $\lambda_{T5}$ coupling since the corresponding interaction term contains the epsilon symbol ($\Gamma^5 \Gamma^{\underline{\mu} \underline{\mu}} = i e_{\underline{\mu} \underline{\nu} \underline{\lambda} \underline{\rho}} \Gamma^{\underline{\lambda} \underline{\rho}}$) and therefore breaks parity. The remaining $\lambda_T$ coupling doesn't introduce any gap in the timelike case and leads to a gap with two nodes in case of spatial condensation of the order parameter.

\section{Numerical treatment}

\subsection*{\label{app:DeTruck}Gauge transformation of the numerical solution}

We obtain the background field profiles by solving the system of nonlinear equations of motion, following from \eqref{equ:Proca_part}, subject to the boundary conditions at the UV boundary \eqref{equ:A_ansatz} and at the horizon \eqref{equ:temperature}. In order to solve this boundary value problem, we rely on 1D relaxation method \cite{Krikun:2018ufr}. For the calculations of the nonlinear backgrounds we use the iterative procedure with Pseudospectral collocation approximations for the derivatives and Chebyshev grids of size $N=60$. For the relaxation to converge, the equations need to have elliptic form, i.e. in 1D contain the second derivative. This is not the case for the Einstein equation, whose $(zz)$ component contains a constraint for $g_{zz}$, instead. We treat this in the standard way using the DeTurck trick \cite{Headrick:2009pv,Wiseman:2011by}, which renders all the equations elliptic. 

The similar subtlety appears in the equations for the Proca field, where the radial component gives an algebraic constrain for $B_z$. The emergent extra gauge symmetry \eqref{equ:Bz_gauge_transform} turns out to be extra handy in this case. By tuning to the ``radial'' gauge $(B_z)_{rad} = 0$ we can get rid of this field component completely. The remaining constraint, following from the $z$-component of the Proca equations of motion, is conserved on the solutions to the remaining equations, therefore it is enough to satisfy it on the boundary. The price to pay for the absence of $B_z$ is the nontrivial profile of the scalar $\Phi$, but the latter has a conventional second order equation of motion and is treated straightforwardly by the numerics. 

In practice, we solve the equations in the ``radial'' gauge and then apply the gauge transformation \eqref{equ:Bz_gauge_transform} with the phase given by the profile of the obtained numerical solution
\begin{equation}
\label{equ:alpha_transform}
\alpha =  \mathrm{arctan} \left(\sqrt{2} \Phi_{rad}/\phi_{rad} \right).
\end{equation}
This leads to a solution in the ``physical'' gauge with $\Phi_{phys} = 0$ and $(B_z)_{phys} = \p_z \alpha/ (\sqrt{2} q)$.

Using the ansatz \eqref{equ:radial_gauge}, \eqref{equ:A_ansatz}, \eqref{equ:ds_ansatz} we introduce the Reissner Nordstr\"om reference metric obtained by setting all ansatz functions to unity in  \eqref{equ:ds_ansatz} and solve for the 8 unknowns $\{A_t(z), B_t(z),T(z), W(z), Z(z), \Phi(z), \phi(z), \xi(z) \}$ in the radial gauge $B_z = 0$. 
The solutions are parametrized by the physically relevant dimensionless parameters $T/\mu$, $p/\mu$, $\lambda/\mu$. We approximating the derivatives with pseudospectral collocation method and use the Chebyshev grid with 60 nodes. This allows us to achieve the temperatures down to reasonably low $T/\mu = 0.001$. 
Given that the symmetry breaking is spontaneous, and we only control the boundary values of the fields as in \eqref{equ:A_ansatz},\eqref{equ:xi_lambda}, setting the sources of the Proca field to zero, there is no algorithmic way of constructing a nontrivial solution in the condensed phase. We use the profiles of the linearized unstable modes at temperatures lower then the critical one in order to seed the condensed solution. 

\begin{figure}
\includegraphics[width=0.45\linewidth]{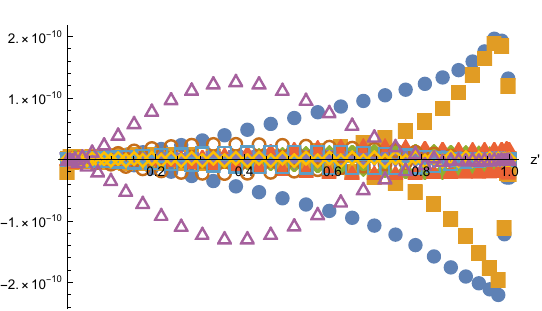}
\includegraphics[width=0.09\linewidth]{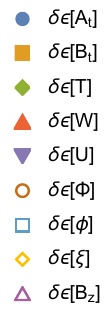}
\includegraphics[width=0.45\linewidth]{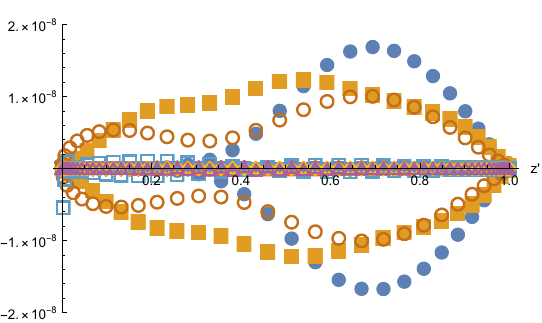}
\caption{\label{fig:satisfy_eoms}Both the numerical solution in the ``radial'' gauge, left panel, and in the one transformed to the ``physical'' gauge, right panel, satisfy the equations of motion.}
\end{figure}

We check, that the numerical gauge transformed with \eqref{equ:alpha_transform} solution still satisfies the numerical equations of motion (as it should, based on analytics). On Fig.\ref{fig:satisfy_eoms} we show on the left panel the values of the equations of motion as evaluated on the original numerical solution in the ``radial'' gauge as well as the same equations evaluated on the transformed solution in the ``physical'' gauge. While some accuracy is clearly lost on the analytical manipulations, the residues on the transformed solutions still have the shape of the numerical noise, meaning that the equations are satisfied.

\subsection*{\label{app:Thermodynamics}Boundary conditions and numerical profiles}

Given that the equations of motion have singular terms at both UV and horizon boundaries, we need to treat boundary behavior of the solutions carefully and adjust the ansatz for the numerical functions we are solving.

The behavior of the solutions near the UV boundary and the horizon can be obtained by expanding the equations at $z \rar 0$ and $z\rar 1$, respectively. For our choice of model parameters \eqref{equ:parameters} 
the expansion at $z\rar 0$ gives
%For the parameters $m=0$, $\phi^{(0)} = \sqrt{3/16}$, $q=2$, $m_{\xi} = -2$
%
\begin{align}
\label{equ:UV_expansions}
A_t &= \mu + \rho z, & T & = 1 - \frac{\lambda^2}{4} z^2 + T^{(3)} z^3,\\
B_t & =z^{-1/2} (1-z) \left(B_t^{(0)} + z^{2}  B_t^{(2)} \right), & W & = 1 - \frac{\lambda^2}{4} z^2 + W^{(3)} z^3,\\
% B_x &= z^{3/2} B_x^{(2)} \\
% B_y &= z^{3/2} B_y^{(2)} \\
% W_y & = 1 - \frac{\lambda^2}{4} z^2 + W_y^{(3)} z^3 \\
\Phi & = z^{-1/2} z^4 \frac{4}{7} \sqrt{3} \mu  B_t^{(2)}, & U & = 1 + \frac{2}{3} \lambda \xi^{(2)} z^3 + U^{(4)} z^4  \\
\phi &= \frac{\sqrt{3}}{4} + \phi^{(3)} z^3, &  \xi & = \lambda z + \xi^{(2)} z^2.
\end{align}
The leading constants correspond to the sources and the subleading to the expectation values. Therefore we identify $\rho$ as the U(1) charge density and we can set $B_t^{(0)}$ to zero for our case of spontaneous symmetry breaking.
For the numerical calculation we factor out the powers of $z$ by reparametrizing the unknown functions as
\begin{align}
\label{equ:numerical_substitutions}
A_t &= \mu (1-z) \tilde{F}_1 (z), & T & = 1 - \frac{\lambda^2}{4} z^2 + z^2 \tilde{F}_3 (z)\\
B_t & =z^{3/2} (1-z) \tilde{F}_2 (z), & W & = 1 - \frac{\lambda^2}{4} z^2 + z^2 \tilde{F}_4 (z)\\
\Phi & = z^{7/2} \tilde{F}_6 (z), & U & = 1 + z^2 \tilde{F}_5(z)\\
\phi &= \frac{\sqrt{3}}{4} + z^2 \tilde{F}_7 (z), & \xi & = z \tilde{F}_8 (z).
\end{align}
The newly introduced functions $\tilde{F}_i$ are regular throughout the bulk and satisfy the simple boundary conditions at the UV boundary.
\begin{equation}
\tilde{F}_1|_{z\rar0} = 1, \qquad \tilde{F}_8|_{z\rar0} = \lambda, \qquad \tilde{F}_{i}|_{z\rar0} \sim O(z), \ i \in (2,7)  
\end{equation}
We can further improve the behavior of the solutions at the horizon by adopting the rescaled radial coordinate \cite{Withers:2014sja,Donos:2014yya}:
\begin{equation}
\label{equ:improved_z}
z = 1 - (1-z')^2.
\end{equation}
Given that $z' \sim \sqrt{f}$ at the horizon, all the regular background profile solutions satisfy the Neuman boundary condition in $z'$. Besides that, the thermal equilibrium requires the metric components to satisfy $U = T$ at the horizon.
\begin{equation}
\p_{z'} \tilde{F}_i|_{z'\rar1} = 0, \qquad \tilde{F}_5|_{z'\rar1} = \tilde{F}_3|_{z'\rar1} - \frac{\lambda^2}{4}
\end{equation}

\begin{figure}
\includegraphics[width=0.45\linewidth]{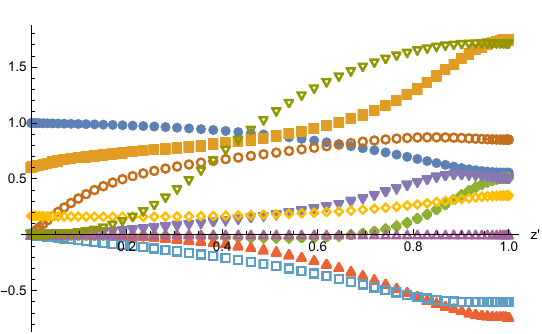}
\includegraphics[width=0.09\linewidth]{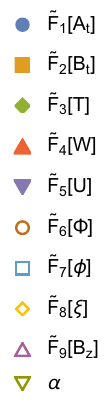}
\includegraphics[width=0.45\linewidth]{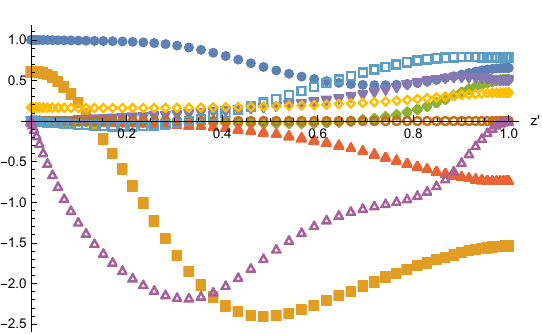}
\caption{\label{fig:numerical_profiles}The solutions to the numerical profiles $\tilde{F}_i[z']$, \eqref{equ:numerical_substitutions} in the ``radial'' gauge (left panel) and the ``physical'' gauge (right panel). The gauge transformation parameter $\alpha(z)$ is shown on the left panel as well.}
\end{figure}

The sample of the numerical profiles, which we end up dealing with, is shown on Fig.\ref{fig:numerical_profiles} (note these examples are obtained on the $n=60$ Chebyshev grid.) Evidently, the profiles of $\tilde{F}_i(z')$ are regular everywhere and behave linearly in the UV.

\subsection*{Thermodynamics}

Following the definition in \cite{Balasubramanian:1999re} and the treatments of \cite{Hartnoll:2007ai, Donos:2012yu,Withers:2013loa}  the boundary stress-energy tensor is
\begin{equation}
T^{b.h.}_{a b} = \frac{1}{8 \pi G} \left[\Theta_{a b} - \Theta \gamma_{a b} - 2 \kappa_1 \frac{2}{L} \gamma_{a b} + 2 \kappa_2 \frac{\lambda^2}{2} \gamma_{a b}  \right],
\end{equation}
where $\Theta$ is the extrinsic curvature at the AdS confomral boundary $z=0$ (with unit normal vector $\hat{n}$ and covariant derivative according to the bulk metric)
\begin{equation}
\Theta^{\mu \nu} = -\frac{1}{2} (\nabla^{\mu} \hat{n}^{\nu} + \nabla^{\nu} \hat{n}^{\mu} ), \qquad \hat{n} = \frac{1}{\sqrt{g_{zz}}} dz, \qquad \Theta \equiv \Theta_{\mu}^{\mu},
\end{equation}
$\gamma_{ab}$ is the pull back of the bulk metric on the boundary and $\lambda$ is the source of the Q-lattice scalar. 
%
% \begin{equation}
% \Theta_{a b} = -\frac{1}{2} \sqrt{g^{zz}} 
% \begin{pmatrix}
% \p_{z} g_{tt} & 0 & 0 \\
% 0 &  \p_z g_{xx} & 0 \\
% 0 & 0 & \p_z g_{yy}
% \end{pmatrix}, \qquad a,b \in \{t,x,y\}
% \end{equation}
% The boundary stress energy tensor reads \cite{Balasubramanian:1999re} and must scale as $z$ for finite energy solutions
% \begin{equation}
% T^{b.h.}_{a b} = \frac{1}{8 \pi G} \left[\Theta_{a b} - \Theta \gamma_{a b} - 2 \kappa_1 \frac{2}{L} \gamma_{a b} + 2 \kappa_2 \frac{ \xi^2}{2} \gamma_{a b}  \right]
% \end{equation}
% Note that in our chosen units $16 \pi G = 1$.
The $\kappa_1$ counter term has been introduced in \cite{Balasubramanian:1999re} and aims at cancelling the divergence of stress energy tensor due to finite curvature, while the term $\kappa_2$ is necessary to cancel the divergent contributions from the Q-lattice, coming from the $z^2$ terms in the UV expansions of the metric profiles \eqref{equ:numerical_substitutions}. Note that no extra counterterms are needed for $\Phi$, $\phi$ or $B_t$ fields, since the corresponding contributions to the on-shell action fall off quickly at $z\rar 0$. 

Given the metric ansatz \eqref{equ:ds_ansatz} and the UV expansions \eqref{equ:UV_expansions}, the stress-energy tensor reads
\begin{gather}
T^{b.h.}_{tt} = 4 \frac{1 - \kappa_1}{z^2} + (\kappa_1 - \kappa_2) \lambda^2 + z \left(-6 + 4 \kappa_1 + \frac{\mu_1^2}{2}(2 \kappa_1 - 3) + (7-4 \kappa_1) T^{(3)} + \frac{2}{3} \lambda \xi^{(2)} (7 - 3 \kappa_2) \right) \\
T^{b.h.}_{xx} = 4 \frac{1 - \kappa_1}{z^2} - (\kappa_1 - \kappa_2) \lambda^2 + z \left(-1 + \frac{\mu_1^2}{4} + (4 \kappa_1 - 7) W_x^{(3)} + \frac{2}{3} \lambda \xi^{(2)} (3 \kappa_2 - 7) \right) 
\\
T^{b.h.}_{yy} = 4 \frac{1 - \kappa_1}{z^2} - (\kappa_1 - \kappa_2) \lambda^2 + z \left(-1 + \frac{\mu_1^2}{4} + (4 \kappa_1 - 7) W_y^{(3)} + \frac{2}{3} \lambda \xi^{(2)} (3 \kappa_2 - 7) \right) 
\end{gather}
Evidently, the divergences cancel after we remove the bookkeeping factors $\kappa_1 = \kappa_2 = 1$. Introducing the appropriate rescaling factors at the boundary we finally get for the stress tensor of the solution
\begin{gather}
T^{CFT}_{tt} = -2 - \frac{\mu_1^2}{2} + 3 T^{(3)} + \frac{8}{3} \lambda \xi^{(2)} \\
T^{CFT}_{xx} = -1 - \frac{\mu_1^2}{4} - 3 W^{(3)} - \frac{8}{3} \lambda \xi^{(2)} \\
T^{CFT}_{yy} = -1 - \frac{\mu_1^2}{4} - 3 W^{(3)} - \frac{8}{3} \lambda \xi^{(2)}
\end{gather}
We use these expressions in order to evaluate the thermodynamical potential shown on Fig.\ref{fig:Back_solutions}
\begin{equation}
\Omega= \epsilon - T s - \mu \rho, \qquad \epsilon \equiv T^{CFT}_{tt}.
\end{equation}
The entropy density $s$ is evaluated as the area of the black hole horizon
\begin{equation}
s= 4 \pi W|_{z=1}
\end{equation}

% \begin{align}
% B_t &= z^{-1/2} B_t^{(0)} + z^{3/2} B_t^{(1)} \\
% \phi &= \phi^{(0)} + \phi^{(1)} z^3 \\
% \Phi &\sim z^2 B_{t}
% \end{align}

% For the numerical simulations we use the ansatz
% \begin{align}
% A_t &=  \mu (1-z) F_1(z) \\
% B_t & = z^{3/2} (1-z) F_2(z) \\
% T & = 1 - \frac{\lambda^2}{4} z^2 + z^2 F_3(z) \\
% W & = 1 - \frac{\lambda^2}{4} z^2 + z^2 F_4(z) \\
% U & = 1 + z^2 F_5(z) \\
% \Phi &= z^{7/2} F_6(z) \\
% \phi &= \frac{\sqrt{3}}{4} + z^2 F_7(z) \\
% \xi &= z F_8(z)
% \end{align}

\subsection*{\label{app:transport}Numerical computation of the transport properties}

In order to evaluate the transport properties we need to solve the system of linear ordinary differential equations which govern the modes excited by the electric or heat current \eqref{equ:transport_perturbations}. In order to render the equations elliptic, we fix the DeDonder gauge for metric fields as in \cite{Rangamani:2015hka} as well as Lorentz gauge for gauge field $A_\mu$.
The near horizon behavior of the modes is controlled by the requirement of regularity in infalling Eddington-Finkelstein coordinates \cite{Rangamani:2015hka}. In particular, we define
\begin{equation}
\{\delta g_{tx}, \delta A_x, \delta g_{xz}, \delta B_x, \delta \bar B_x, \delta \chi_1 \} = (1-z^4)^{-\frac{i \omega}{P(1)}}\{\delta \hat g_{tx}, \delta \hat A_x, \frac{\delta \hat g_{xz}}{f(z)}, \delta \hat B_x, \delta \hat{\bar{B_x}}, \delta \hat \chi_1 \},
\end{equation}
where $P(z) \equiv f(z)/(1-z)$.

We solve the linear system of equations by means of discretization on a linearly spaced grids of sizes $N=120 \dots 240$. We obtain the solution by solving the resulting algebraic linear system using Wolfram Mathematica's \texttt{LinearSolve[]} routine \cite{Mathematica}.

For the sake of numerical accuracy it turns out quite useful to separate explicitly the sources and the subleading branches of the perturbative solutions near the UV boundary. By analyzing the equations of motion at $z\rar 0$ we can write down the expansions
\begin{align}
\delta \hat g_{tx} &= \frac{1}{z^2} \delta T_{tx} + \delta T_{tx} \left(- \lambda^2 + \frac{2}{3} \omega^2 \right) + \tilde g_{tx}(z), \qquad
\delta \hat A_x = \delta E_x + \tilde A_x(z),\\
\delta \hat g_{xz} &= \frac{1}{z} \delta T_{tx} \frac{2}{3} i \omega - \delta T_{tx} \frac{1}{3}i \omega + z \delta T_{tx}  \frac{2}{3} i \omega (2 \omega^2 - 3 \lambda^2) + z \tilde g_{xz}(z),\\
\delta \hat B_{x} &= \sqrt{z} \tilde B_{x}(z), \qquad
\delta \hat{\bar{B_{x}}} = \sqrt{z} \tilde{\bar{B_{x}}}(z),\qquad
\delta \hat \chi_1 = \tilde \chi_1.
\end{align}
In this formulation all the boundary sources appear explicitly in the equations and the all the functions with tildas, which we are solving for numerically, have Dirichlet boundary conditions at $z=0$ with their first derivatives corresponding to the subleading coefficient in the solution. This latter coefficient is in its turn related to the expectation value of the corresponding operator, so reading off the first derivatives of the solutions in presence of perturbative sources we obtain the necessary two-point functions. 
\begin{equation}
\la J_x \ra(x) = \p_z \tilde{A}_x(z) \Big|_{z=0}, \qquad \la T_{tx} \ra(x) =- \p_z \tilde{g}_{tx}(z) \Big|_{z=0}.
\end{equation}
In order to get rid of the contribution of the contact terms we subtract the real part of the correlator at zero frequency from the AC results \cite{Kim:2014bza}. As a nontrivial check of our calculations, we make sure that $\la J_x T_{tx} \ra (\omega)$ matches with $\la T_{tx} J_x \ra (\omega)$, leading to $\alpha=\bar \alpha$ in \eqref{eq:sigma_matrix}. Our analysis is analogous to the one described in \cite{Andrade:2022udb}.

In case of $P+iP$ the linear perturbations, which couple to the current in $x$-direction are
\begin{equation}
\label{equ:transport_perturbations_pip}
\{\delta g_{tx}, \delta A_x, \delta g_{xz}, \delta g_{ty}, \delta A_y, \delta g_{yz}, \delta B_{t}, \delta B_{z}, \delta \bar B_{t}, \delta \bar B_{z}, \delta \Phi, \delta \bar \Phi, \delta \chi_1, \delta \chi_2 \} \sim e^{i \omega t}.
\end{equation}
The excitation of transverse current relates to the nonzero Hall conductivity in $P+iP$ state.
Due to the considerably more complex structure of the equations, we had to rise the numerical grid sizes to $N=360$ linear points in order to produce the plots on Figs.\,\ref{fig:sigma}, \ref{fig:kappa}.

\subsection*{\label{app:fermion}Numerical computation of the fermionic spectrum}
The Dirac equation is linear and therefore it can be solved in one step by inverting the system of linear equations on the computational grid. We use the linear spaced grids of the sizes $N=60 \dots 240$ and interpolate the backgrounds solutions on them. The solutions are obtained by applying Wolfram Mathematica \texttt{LinearSolve[]} routine to the discretized equations. 

The rescaled radial coordinate \eqref{equ:improved_z} fits perfectly to the fermionic problem, since the solutions approach horizon as $\sqrt{f}$. The couplings \eqref{equ:fermion_interaction_terms} do not affect either the near of the near boundary behavior of the fields. We use the ansatz
\begin{equation}
\zeta_i \sim e^{i (- \omega t + k_x x + k_y y)}
\end{equation}
for the fermionic components, taking into account momenta in both directions. However, all the backgrounds considered in this work are isotropic, so turning on $k_x$ or $k_y$ momenta and matching the results allows us to check that our numeric algorithm is consistent. In all regards we follow the standard treatment described i.e. in \cite{Balm:2019dxk,Rodgers:2022uvs}.

\section{\label{app:Green_functions}Perturbative treatment of the Meissner effect}

In this appendix we develop the Green's function machinery for the perturbative treatment of the Meissner effect in holographic superconductor. We start by reviewing the case of a scalar superconductor from \cite{Natsuume:2022kic} and then perform the calculation in the timelike case.

As discussed in the main text we will consider the regime where the order parameter is small and neglect the backrection on the geomtry, therefore the metric used in all calculations in this section is given by the Schwarzschild-AdS$_{4}$ black hole

\begin{eqnarray}
ds^{4} 
%&=& r^2 (-f dt^{2} + dx^{2} + dy^{2}) + \frac{dr^{2}}{r^{2}f} \\
&=& \left( \frac{1}{z} \right) ^{2} (-f dt^{2} + dx^{2} + dy^{2}) + \frac{dz^{2}}{z^{2} f}, \\
f &=& 
%1- \left( \frac{r_{0}}{r} \right) ^{3} = 
1-z^{3},
\end{eqnarray}
where we set the AdS curvature radius to 1 and rescale the radial coordinate so that the horizon is located at $z_h = 1$.

\subsection*{Scalar case}
We warm up by considering the Meissner effect in an holographic superconductor with a scalar condensate $\psi$
Applying the magnetic field perturbatively gives us a linearized bulk equation that, after making a Fourier transformation, reads
\begin{equation}
\label{0.1}
\partial_{z} \left( f A_{y} \right) - k^{2} A_{y} - \frac{2 q_{\psi}^{2} A_{y} \psi^{2}}{z^{2}}.
\end{equation}
Assuming $\psi \sim \epsilon \rar 0$ we perform the expansion over $k$ and $\epsilon$ of the form
\begin{eqnarray}
A_{y} 
%&=& \sum_{j} \sum_{i} q_{f}^{2i} q^{2j} A_{yij}   \\
\label{0.1.0}
&=& A_{y00} + k^{2} A_{y01} + \epsilon^2 q_{\psi}^{2} A_{y10} + \cdots,
\end{eqnarray}
which after substituing in \ref{0.1}, leads to a set of equations 
\begin{eqnarray}
\label{0.2}
 \partial_{z} \left( f \partial_{z} A_{y00} \right) &= 0 \\
 \partial_{z} \left( f \partial_{z} A_{y01} \right) &= A_{y00} \\
 \partial_{z} \left( f \partial_{z} A_{y10} \right) &= \frac{2 \psi^{2} A_{y00}}{z^{2}}.
\end{eqnarray} 
We start by solving the equation for the leading order solution $A_{y00}$. The general solution to the linear equation \eqref{0.2} can be constructed using the linearly independent basis functions 
\begin{equation}
\label{equ:JBA_basis}
J_A(z') = 1, \qquad K_A(z) = \int_{0}^{z} ds \frac{1}{f(s)}, \qquad \p_z  K_A(z) \Big|_{z\rar 0} = \frac{1}{f(z)}\Big|_{z\rar 0}  = 1
\end{equation}
which we define using the boundary conditions
\begin{equation}
\label{J_A_bc}
J_A(z) \Big|_{z\rar 1} \sim \mathrm{finite}, \qquad K_A(z) \Big|_{z\rar 0} \rar 0.
\end{equation}
The leading order solution to \eqref{0.2} is the bulk-to-boundary propagator satisfying \eqref{equ:bulk_to_boudary_prop}, $A_{y00}(z)|_{z\rar 0} = 1$, which leads to a simple solution
\begin{equation}
\label{equ:leading_A}
A_{y00}=1.
\end{equation}

In order to obtain the solutions at the subleading order, we employ the Green's function method. We consider the Green's function satisfying the equation
\begin{equation}
 \partial_{z} \left( f \partial_{z} G_A(z,z') \right) = \delta(z-z').
\end{equation}
The solution can be represented in the form
\begin{eqnarray}
G_{A}(z,z') 
%&=& \left\lbrace \begin{array}{c} \alpha_{1}(s) J_{A} (u) + \alpha_{2}(s) K_{A}(u); u > s \\ \alpha_{3}(s) J_{A} (u) + \alpha_{4}(s) K_{A}(u); u<s \end{array}\right. \nonumber \\
&=& \left\lbrace \begin{array}{c} \alpha_{1}(z') J_{A} (z); z > z' \\ \alpha_{2}(z') K_{A} (z); z<z' \end{array}\right. , 
\end{eqnarray}
where we use the boundary conditions $G_{A}(0,s)=0$, $G_{A}(1,s) < \infty$. In order to obtain $\alpha_{1}$ and $\alpha_{2}$, we use the continuity conditions
\begin{eqnarray}
& G_{A}(z' + \epsilon , z')=G_{A}(z' - \epsilon , z') \\
& \int_{z' - \epsilon}^{z' + \epsilon} dz \partial_{z} \left[ f(z) \partial_{z} \left( G_{A}(z,z') \right) \right] = 1,
\end{eqnarray} 
from which we can write it in the form
\begin{eqnarray}
G_{A}(z,z') &=& \frac{1}{f(z') (K_A(z')J_A'(z') - J_A(z')K_A'(z'))} \left\lbrace \begin{array}{c} K_{A} (z') J_{A} (z); z > z' \\ J_{A} (z') K_{A} (z); z<z' \end{array}\right.
\end{eqnarray}
Noteworthy, the denominator here is the Wronskian
\begin{align}
W(z') \equiv K_A(z') J_A'(z') - J_A(z') K_A'(z') & = C_A \exp\left(- \int \limits_{0}^{z'} ds \frac{f'(s)}{f(s)} \right) 
%\\
%& = C_B \exp\left(- \log(f(z')) + \log(f(0))\right) 
= \frac{C_A}{f(z')}
\end{align}
and the constant $C_A$ can be evaluated by considering the basis solution at any convienient point, evaluating the Wronskian at $z'=0$ we obtain using \eqref{equ:JBA_basis}
\begin{align}
\label{equ:CA}
C_A = f(0) \left[K_A(0) J_A'(0) - J_A(0) K_A'(0)\right] &= -1.
%K_A(z') J_A'(z') - J_A(z') K_A'(z') & = \frac{C_A}{f(z')} \\
%- K_A'(z') & = \frac{C_A}{f(z')} \\
%- \frac{1}{f(z')} & = \frac{C_A}{f(z')} \\
%C_A &= -1
\end{align}
This finally gives us the Green's function
\begin{equation}
\label{equ:Green_A}
G_{A}(z,z') =\frac{1}{C_A} \left\lbrace \begin{array}{c} K_{A} (z') J_{A} (z); z > z' \\ J_{A} (z') K_{A} (z); z<z' \end{array}\right..
\end{equation}
Making use of \eqref{equ:Green_A} and \eqref{equ:leading_A} we can now evaluate the corrections to the leading solution:
\begin{eqnarray}
\label{equ:scalar_perturbation_solutions}
&A_{y01}(z)= \int_{0}^{1} dz' G_{A}(z,z') 1  \\
&A_{y10}(z)= \int_{0}^{1} dz' G_{A}(z,z') \left(  \frac{2 \psi^{2}(z')}{z'^{2}} \right) . 
\end{eqnarray}
Given these expressions, we can evaluate the subleading contribution to the expectation value of the current in \eqref{equ:semiclassical_maxwell}: $\langle \tilde{\mathcal{J}}_{y} \rangle = \partial_{z} \tilde{A}_{y} \vert_{z=0}$. Note that since we are considering the asymptote $z\rar0$, then only the $z'>z$ branch of the Green's function will contribute to the finite integral in \eqref{equ:scalar_perturbation_solutions}.
\begin{eqnarray}
\partial_{z} A_{y01} \vert_{z=0} &=& 
%\int_{0}^{1} ds \partial_{u} \left( G_{A}(u,s) \right) \vert_{u=0} \nonumber  \\ &=& 
\frac{1}{C_A} \left[ \int_{z}^{1} dz' J_{A}(z') \left( \partial_{z} K_{A}(z) \right) \right] \vert_{z=0} \nonumber 
%\\
%&=& \frac{1}{f(s) (C_{A} e^{- \int_{0}^{s} du' \frac{f'}{f}})} \left[ \int_{0}^{1} ds \left( \frac{1}{f(0)} \right) \right] 
= \frac{1}{C_{A}} 
\end{eqnarray}
and
\begin{eqnarray}
\partial_{z} A_{y10} \vert_{z=0} &=& 
%\int_{0}^{1} ds \partial_{u} \left( G_{A}(u,s) \right) \left(  \frac{2 \psi^{2}(s)}{s^{2}} \right) \vert_{u=0} \nonumber \\
%&=& 
%\frac{1}{C_A} \left[ \int_{z}^{1} dz' J_{A}(z') \left( \partial_{z} K_{A}(z) \right) \right] \left(  \frac{2 \psi^{2}(z')}{z'^{2}} \right) \vert_{z=0} 
%\nonumber 
%\\
%&=& \frac{1}{f(s) (C_{A} e^{- \int_{0}^{s} du' \frac{f'}{f}})} \left[ \int_{0}^{1} ds \left( \frac{1}{f(0)} \right) \right] \left(  \frac{2 \psi^{2}(s)}{s^{2}} \right) = 
%=
\frac{1}{C_{A}} \int_{0}^{1} dz' \left(  \frac{2 \psi^{2}(z')}{z'^{2}} \right).
\end{eqnarray}

Finally, given the expansion \ref{0.1.0} and recalling $C_A = -1$ we obtain the current as
\begin{eqnarray}
\label{0.3}
& \partial_{z} \tilde{A}_{y} \vert_{z=0} = - k^{2} \mathcal{A}_y(k) - 2 I \partial_{z} \mathcal{A}_y(k) + \cdots  \qquad 
I= \int_{0}^{1} ds \left(  \frac{ \psi^{2}(s)}{s^{2}} \right),
\end{eqnarray}
This result is equivalent to the one in \cite{Natsuume:2022kic}.

\subsection*{Timelike case}

In order to evaluate the magnetic penetration length in the timelike superconductor, follow the same strategy, turning on the perturbative $\delta A_y$ component, but in this case it couples linearly with the $\delta B_y$ component of the charged vector, therefore instead of \eqref{0.1} we have to develop a perturbative solution to the system of equations \eqref{equ:dAy_equations_timelike}.
% \begin{align}
% \p_z \left(f \p_z \delta \tilde{A}_y  \right) - & k^2 \delta \tilde{A}_y + \frac{2 q_B^2}{f} \left(B_t^2 - f^2 B_z^2 \right) \delta \tilde{A}_y \\
% + & 4 q_B f B_z \, \p_z \delta B_y^R + 2 q_B \left( \p_z (f B_z) - \frac{q_B}{f} A_t B_t \right) \delta B_y^R = 0
% \\
% \p_z \left(f \p_z \delta \tilde{B}_y^R  \right) - & k^2 \delta \tilde{B}^R_y - \frac{q_B^2 \phi^2}{z^2} \tilde{B}^R_y + \frac{q_B^2}{f} A_t^2 \delta \tilde{B}_y^R - 2 q_B^2 B_z^2 \delta \tilde{B}_y^R \\
% - & 2 f q_B B_z \p_z \delta \tilde{A}_y - q_B \left( \p_z (f B_z) + \frac{q_B}{f} A_t B_t \right)  \delta \tilde{A}_y=0
% \end{align}

Expanding these equations in the strength of the condensate ($B_t \sim B_z \sim \epsilon$), we get order-by-order (we focus on $k=0$ case here, since this correction is identical to the scalar case)
\begin{align}
\p_z \left(f \p_z \delta \tilde{A}_y^{0}  \right) =& 0 \\
\notag
\p_z \left(f \p_z \delta \tilde{A}_y^{2}  \right) =& -\frac{2 q_B^2}{f} \left(B_t^2 - f^2 B_z^2 \right) \delta \tilde{A}_y \\
\notag
& - 4 q_B f B_z \, \p_z \delta B_y^{1} - 2 q_B \left( \p_z (f B_z) - \frac{q_B}{f} A_t B_t \right) \delta B_y^{1} 
\\
\notag
\p_z \left(f \p_z \delta \tilde{B}_y^{1} \right) & - q_B^2 \left(\frac{\phi^2}{z^2} - \frac{A_t^2}{f} \right) \delta \tilde{B}^{1}_y = \\
\notag
=& 2 q_B f B_z \p_z \delta \tilde{A}_y^{0} + q_B \left( \p_z (f B_z) + \frac{q_B}{f} A_t B_t \right)  \delta \tilde{A}^{0}_y
\end{align}

Again, the leading order solution is $\delta A_y^0 = 1$. For $B$-field it is and $\delta B_y^0 = 0$, since there are no sources, no finite boundary conditions for it.
We start by solving for $\delta \tilde{B}^{1}_y$. The corresponding Green's function satisfies the equation
\begin{equation}
\label{equ:Green_B}
\p_z \left(f \p_z G_B(z,z') \right) - q_B^2 \left(\frac{\phi^2}{z^2} - \frac{A_t^2}{f} \right) G_B(z,z') = \delta(z-z')
\end{equation}
and can be constructed out of the basis functions with particular boundary conditions:
\begin{equation}
J_B(z) \Big|_{z\rar 1} \sim \mathrm{finite}, \qquad K_B(z) \Big|_{z\rar 0} \rar 0.
\end{equation}
Unlike the case of the $A$-field the basis functions have to be obtained solved numerically. The equation for $B$-field Green's function \eqref{equ:Green_B} contains the background profiles of $\phi$ and $A_t$. Within our approximation of the small infinitesimal order parameter, however, we can consistently use normal state profiles for this calculation
\begin{equation}
\phi = \frac{\sqrt{3}}{2 q_B}, \qquad A_t = \mu (1-z).
\end{equation}
In complete analogy with the $\delta A$ case the Green's function of the $B$-field can be represented as
%
\begin{comment}
The Green's function is constructed as
\begin{align}
G_B(z,z') = 
M(z')
\left\{
\begin{matrix}
\alpha_1 (z') K_B(z) , \quad z < z' \\
\alpha_2 (z') J_B(z) , \quad z > z' \\
\end{matrix}
\right.
\end{align}
The coefficients $\alpha_1, \alpha_2$ are fixed by requirement of continuity:
\begin{align}
G(z,z')\Big|_{z \rar z' - 0} &= G(z,z')\Big|_{z \rar z' + 0}
\alpha_1 (z') K_B(z') &= \alpha_2 (z') J_B(z')
\end{align}
Which means $\alpha_1 = J_B(z'), \alpha_2 = K_B(z')$.

On the other hand the overall coefficient $M$ is fixed by integrating the equation \eqref{equ:Green_B} from $z=z'-0$ to $z=z'+0$ over the delta function.

\begin{align}
\p_z G_B(z,z') \Big|_{z=z' + 0} -  \p_z G_B(z,z') \Big|_{z=z' - 0}   &= \frac{1}{f(z')} \\
M(z') \left( K_B(z') J_B'(z') - J_B(z') K_B'(z') \right) &= \frac{1}{f(z')} \\
M(z') =  \frac{1}{f(z')} \left( K_B(z') J_B'(z') - J_B(z') K_B'(z') \right)^{-1}
\end{align}

We can recognize the Wronskian and employ the theorem
\begin{align}
W(z') \equiv K_B(z') J_B'(z') - J_B(z') K_B'(z') & = C_B \exp\left(- \int \limits_{0}^{z'} ds \frac{f'(s)}{f(s)} \right) \\
& = C_B \exp\left(- \log(f(z')) + \log(f(0))\right) = \frac{C_B}{f(z')}
\end{align}

This gives us for the $M(z')$ coefficient
\begin{align}
M(z') =  C_B^{-1}
\end{align}

\end{comment}
%
%And the Green's function reads
\begin{align}
G_B(z,z') = 
\frac{1}{C_B}
\left\{
\begin{matrix}
J_B (z') K_B(z) , \quad z < z' \\
K_B (z') J_B(z) , \quad z > z' \\
\end{matrix}
\right.
\end{align}
with the constant $C_B$ defined as
\begin{equation}
C_B = f(z') \left( K_B(z') J_B'(z') - J_B(z') K_B'(z') \right)
\end{equation}
for any convenient $z'$.

We use this Green's function in order to obtain $\delta B_y^1$:
\begin{align}
\delta B_y^1(z') &= \int \limits_{0}^{1} ds G_B(z',s)  q_B \left( \p_{s} (f(s) B_z(s)) + \frac{q_B}{f(s)} A_t(s) B_t(s) \right) \\
%
%&=  J_B(z') \int \limits_{0}^{z'} ds K_B (s) q_B \left( \p_{s} (f(s) B_z(s)) + \frac{q_B}{f(s)} A_t(s) B_t(s) \right) \\
%&+
%K_B(z') \int \limits_{z'}^{1} ds J_B (s)  q_B \left( \p_{s} (f(s) B_z(s)) + \frac{q_B}{f(s)} A_t(s) B_t(s) \right) \\
\notag
&=  \frac{J_B(z')}{C_B} \int \limits_{0}^{z'} ds K_B (s) \mathcal{H}(s) +
\frac{K_B(z')}{C_B} \int \limits_{z'}^{1} ds J_B (s) \mathcal{H}(s) \\
\mathcal{H}(s) &\equiv  q_B \left( \p_{s} (f(s) B_z(s)) + \frac{q_B}{f(s)} A_t(s) B_t(s) \right)
\end{align}

As a next step, we compute $\delta A_y^{(2)}$. The Green's function is the same as in the scalar case discussed above \eqref{equ:Green_A}.
%
% \begin{align}
% G_A(z,z') = 
% \frac{1}{C_A}
% \left\{
% \begin{matrix}
% J_A (z') K_A(z) , \quad z < z' \\
% K_A (z') J_A(z) , \quad z > z' \\
% \end{matrix}
% \right.
% \end{align}
%
% Note that the basis functions for $A$ field are very simple:
% \begin{equation}
% J_A(z') = 1, \qquad K_A(z) = \int_{0}^{z} ds \frac{1}{f(s)}, \qquad \p_z  K_A(z) \Big|_{z\rar 0} = \frac{1}{f(z)}\Big|_{z\rar 0}  = 1
% \end{equation}
% Moreover the constant $C_A$ can be evaluated as
% \begin{align}
% K_A(z') J_A'(z') - J_A(z') K_A'(z') & = \frac{C_A}{f(z')} \\
% - K_A'(z') & = \frac{C_A}{f(z')} \\
% - \frac{1}{f(z')} & = \frac{C_A}{f(z')} \\
% C_A &= -1
% \end{align}
%
The solution is
\begin{align}
\delta A^2_y(z) &= \int \limits_{0}^{1} dz' G_A(z,z') \frac{2 q_B^2}{f(z')} \left(- B_t(z')^2 + f(z')^2 B_z(z')^2 \right) \\
\notag
& - \int \limits_{0}^{1} dz' G_A(z,z') 4 q_B f(z') B_z(z') \, \p_z' \delta B_y^{1}(z') \\
\notag
& - \int \limits_{0}^{1} dz' G_A(z,z') 2 q_B \left( \p_z' (f(z') B_z(z')) - \frac{q_B}{f(z)} A_t(z') B_t(z') \right) \delta B_y^{1}(z'). 
\end{align}
We can immediately calculate the current, by evaluating $\p_z \delta A^2_y$ at $z \rar 0$. Again in the integrals the $z' > z$ region will give the dominant contribution and we can substitute $G_A(z,z') = J_A(z') K_A(z)$.
We get
\begin{align}
\label{equ:A2_solution}
\p_z \delta A^2_y(z) \Big|_{z\rar 0} &= \frac{1}{C_A} \int \limits_{0}^{1} dz' \frac{2 q_B^2}{f(z')} \left(- B_t(z')^2 + f(z')^2 B_z(z')^2 \right) \\
\notag
& - \frac{1}{C_A} \int \limits_{0}^{1} dz' 4 q_B f(z') B_z(z') \, \p_{z'} \delta B_y^{1}(z') \\
\notag
& - \frac{1}{C_A} \int \limits_{0}^{1} dz' 2 q_B \left( \p_{z'} (f(z') B_z(z')) - \frac{q_B}{f(z)} A_t(z') B_t(z') \right) \delta B_y^{1}(z') 
\end{align}
It is useful to take the integral by parts and transform the second line, getting rid of the derivative of $\delta B_y^1$:
\begin{multline}
\int \limits_{0}^{1} dz' 4 q_B f(z') B_z(z') \, \p_{z'} \delta B_y^{1}(z') = \\
4 q_B f(z') B_z(z') \, \delta B_y^{1}(z') \bigg|_{0}^{1} - \int \limits_{0}^{1} dz' 4 q_B \p_z' (f(z') B_z(z') ) \, \delta B_y^{1}(z')
\end{multline}
The boundary values of the first part are both zero due to $B_z (0) =0$ and $f(1) = 0$ and the last term transforms the third line in \eqref{equ:A2_solution}, which assumes the shape (We substitute $C_A = -1$)
\begin{align}
\p_z \delta A^2_y(z) \Big|_{z\rar 0} &= -  \int \limits_{0}^{1} dz' \frac{2 q_B^2}{f(z')} \left(- B_t(z')^2 + f(z')^2 B_z(z')^2 \right)  - \int \limits_{0}^{1} dz' 2 \mathcal{H}(z') \delta B_y^{1}(z') 
\end{align}

Finally, substituting the expression for $\delta B_y^{1}(z')$ we get
\begin{align}
\p_z \delta A^2_y(z) \Big|_{z\rar 0} &= - 2 \int \limits_{0}^{1} dz' \frac{q_B^2}{f(z')} \left(- B_t(z')^2 + f(z')^2 B_z(z')^2 \right) \\
\notag
& - 2 \int \limits_{0}^{1} dz' \mathcal{H}(z') \left[\frac{J_B(z')}{C_B} \int \limits_{0}^{z'} ds K_B (s) \mathcal{H}(s) +  \frac{K_B(z')}{C_B} \int \limits_{z'}^{1} ds J_B (s) \mathcal{H}(s) \right] \\
\notag
\mathcal{H}(s) &\equiv  q_B \left( \p_{s} (f(s) B_z(s)) + \frac{q_B}{f(s)} A_t(s) B_t(s) \right)
\end{align}

\bibliographystyle{JHEP-2}
\bibliography{f-odd}

\end{document}